# Fracture of the $C15$ $CaAl_2$ Laves phase at small length-scales

James P. Best[*,†], Anwesha Kanjilal[†], Alireza Ghafarollahi, Uzair Rehman, Chunhua Tian[a)], Hanna Bishara[b)], M. Kamran Bhat, Leon Christiansen, Erik Bitzek, Frank Stein, Gerhard Dehm[*]

*Max-Planck-Institut für Eisenforschung GmbH, Max-Planck-Str. 1, D-40237 Düsseldorf, Germany*

[a)] *Present address: Laboratory for Mechanics of Materials and Nanostructures, Empa – Swiss Federal Institute for Materials Science and Technology, CH-3603 Thun, Switzerland*

[b)] *Present address: Department of Material Science and Engineering, Tel Aviv University, Ramat Aviv 6997801, Tel Aviv, Israel*

[†] *These authors contributed equally to this work*

* *Corresponding authors: j.best@mpie.de and g.dehm@mpie.de*

## Abstract

The cubic $C15$ $CaAl_2$ Laves phase is an important brittle intermetallic precipitate in ternary Mg-Al-Ca structural alloys. Although knowledge of the mechanical properties of the coexisting phases is essential for the design of improved alloys, the fracture toughness has not yet been studied experimentally due to the need for miniaturised testing methods. Here, micropillar splitting and microcantilever bending methods are used to experimentally determine the fracture toughness of the $CaAl_2$ Laves phase. It is found that the toughness value of ~1 MPa·√m from pillar splitting with a sharp cube corner geometry is largely insensitive to sample heat treatment, the ion beam used during fabrication, micropillar diameter, and surface orientation. From correlative nanoindentation and electron channelling contrast imaging supported by electron backscatter diffraction, fracture is observed to take place mostly on {011} planes. Atomistic fracture simulations on a model $C15$ Laves phase showed that the preference of {011} cleavage planes over the more energetically favourable {111} planes is due to lattice trapping and kinetics controlling fracture planes in complex crystal structures. Using rectangular microcantilever bending tests where the notch plane was misoriented to the closest possible {112} cleavage plane by ~8°, and the closest {001}, {011} and {111} plane by >20°, a toughness of ~2 MPa·√m was determined along with the electron microscopy observation of significant deviations of the crack path, demonstrating that preferential crystallographic cleavage planes determine the toughness in this material. Further investigation using pentagonal microcantilevers with precise alignment of the notch with the cleavage planes revealed similar fracture toughness values for different low-index planes. The results presented here are the first detailed experimental study of fracture toughness of the *C*15 $CaAl_2$ Laves phase, and can be understood in terms of crack plane and crack front dependent fracture toughness.

## 1. Introduction

Laves phases are among the most common intermetallic phases and are frequently present in metallic alloys as nano- to microscale inclusions. These tetrahedrally close-packed intermetallic phases either have a cubic (*C*15, $MgCu_2$ type) or hexagonal crystal structure (*C*14, $MgZn_2$ type, or *C*36, $MgNi_2$ type) [1]. They are generally hard and brittle, and therefore their presence largely determines the response of engineering materials to deformation [2]. In aluminium and magnesium alloys, the cubic *C*15 $CaAl_2$ Laves phase is often present. In the Mg-based alloy AZ31, for example, the $CaAl_2$ Laves phase has been demonstrated to promote grain refinement during alloy processing [3]. The precipitation of Laves phases can be favourable to improve the strength of Mg alloys [4]. In particular, $CaAl_2$ is known to be deformable, and precipitation of this phase is often considered attractive due to the action of the inclusions in impeding dislocation motion leading to strong hardening [4, 5]. This is of particular importance for highly formable Mg-Al-Ca alloys, which currently attract attention due to the non-rare-earth combination of Al and Ca promoting ⟨*c*+*a*⟩ dislocation plasticity in Mg when in solid solution, while suppressing twin formation [6]. $CaAl_2$ precipitates have recently been shown to plastically deform via full and partial dislocations (the latter leaving stacking faults), relieving interfacial stress concentrations at the Mg matrix and improving tensile elongation in a Mg-Al-Ca alloy [4].

Currently, few studies report on the room temperature deformation behaviour of Laves phases [7]. As most readily available as precipitates on nano- to micrometre length scales, there is an inherent difficulty in extracting their intrinsic mechanical properties from these inclusions. For determination of the critical resolved shear stress (CRSS) of $CaAl_2$, microcompression approaches harnessing focussed ion beam milling of micropillars have recently been extended to precipitates in a Mg-Al-Ca alloy [5], where a CRSS of 164.9 MPa was determined for the {111}⟨1-10⟩ slip system. For determination of the fracture toughness, however, larger specimens >10 μm are generally required to accommodate the processing of a pre-fabricated defect (usually a sharp notch) and the formation of a $K_I$-dominated stress field under the defect; dimensions which far exceed the geometry of intermetallic inclusions. Moreover, sub-micron volumes of brittle material generally favour plastic deformation [8], suppressing an investigation of fracture.

A promising approach is to investigate bulk intermetallic samples. However, for fracture toughness determination, testing of bulk-scale Laves phases is precluded by significant concentrations of pre-existing cracks and defects. This makes classical macroscopic fracture toughness testing of a single-phase Laves phase sample practically impossible, since the production of defect-free single-phase material on a centimetre scale already poses a significant challenge [9]. Nevertheless, it is possible to comprehensively investigate the mechanical response of Laves phases by applying small-scale testing methods in extended, single Laves phase grains in bulk pieces. In studying plasticity, Zehnder *et al.* investigated the *C*14 $CaMg_2$ Laves phase through micropillar compression [10], which was recently extended to compression to 250 °C by Freund *et al.* on the same bulk *C*14 sample [11]. Furthermore, similar micropillar compression studies were recently performed on bulk single-phase *C*15 $CaAl_2$ Laves phase specimens [12]. Bulk-scale diffusion couples are also a viable approach for studying the mechanical properties of Laves phases, including as a function of local chemistry. This was demonstrated by Luo *et al.* through nanoindentation and micropillar compression tests on the binary $NbCo_2$ Laves phase, which exists in the three crystallographic structure variants *C*14, *C*15, and *C*36 depending on composition [13, 14]. By a similar approach, the fracture toughness was also determined for the $NbCo_2$ Laves phases using focussed ion beam (FIB) milled microcantilevers [15]. In that work, the authors determined that local composition affected the elastic modulus and hardness, but a constant fracture toughness value of ~4.2 MPa·√m was extracted for all three Laves phases irrespective of crystal structure and stoichiometry, with the notch plane for the *C*15 phase aligned along close-packed {111} planes.

As the fracture toughness of the *C*15 $CaAl_2$ phase has not yet been studied experimentally, a comprehensive micromechanical toughness study is presented here. A heat-treated cast material containing large grains of the $CaAl_2$ Laves phase was fabricated and the fracture toughness was investigated using both micropillar splitting and microcantilever testing techniques. Spherical nanoindentation allowed for a rationalisation of preferential fracture planes in this system, which was further supported by atomistic simulations on a model *C*15 Laves phase.

## 2. Materials and methods

### *2.1. Sample fabrication and surface characterisation*

The intermetallic *C*15 $CaAl_2$ specimen was cast into a cylindrical mould (size D15) to target a composition of 67 at.% Al and 33 at.% Ca. Granular elemental materials for Al (99.999% purity, HMW Hauner GmbH & Co. KG, Germany) and Ca (>98.8% purity, Alfa Aesar, Germany – Lot no: 61200136) were used in sample preparation. The sample was then annealed at 600°C under an argon atmosphere for 24 hours to improve homogeneity, reduce residual stresses, and promote the target phase growth. Subsequently, the annealed material was cut into smaller pieces using electric discharge machining, and further analysis presented in this manuscript is for the cross-section of the cast cylindrical rod. Details of the sample surface preparation can be found in Ref. [16]. Imaging of the sample was performed in a Zeiss Auriga scanning electron microscope (SEM) at 10 kV primarily using both backscattered electron (BSE) and secondary electron (SE) detectors, while electron backscatter diffraction (EBSD) measurements were conducted using the same machine at 20 kV and a step size of 0.3 µm with an EDAX TSL-OIM version 6.0 data acquisition system. The chemical composition was analysed using electron probe microanalysis (EPMA) on the metallographically prepared specimen. EPMA analysis was performed at an acceleration voltage of 10 kV and the resulting chemical composition was averaged over 12 areas of interest per phase.

### *2.2. Nanoindentation*

Nanoindentation using a commercial nanoindenter (G200, Brucker/KLA) was performed to determine the hardness and elastic modulus of the heat-treated intermetallic phase. A diamond Berkovich tip (Synton-MDP, Switzerland) was used with a constant indentation strain rate of 0.05 $s^{-1}$, maximum indentation depth 500 nm and peak hold time of 2 seconds. The elastic modulus $E$ and hardness $H$ were determined using the Oliver-Pharr method [17]. Before performing the test, tip and instrument calibrations were performed on a fused silica reference sample. Spherical nanoindentation was performed using a diamond spherical counterbody with radius 1 µm (Synton-MDP, Switzerland) into three grains of the annealed sample with distinct surface crystallographic planes, to a displacement target set-point of 500 nm. In each studied grain, 5 indents were performed using the G200 nanoindenter. The deformation zones, and particularly cracks around the indents, were characterised in a SEM (Zeiss MERLIN), by electron channelling contrast imaging (ECCI), using an accelerating voltage of 30 kV, a beam current of 2 nA, and a working distance of 7.5 mm. For the crack trace analysis, EBSD in conjunction with the ECCI were employed to reveal the crystallographic family of crack planes. The asymmetric domain method was employed to obtain Euler angles in each grain reside in the same symmetric subset of orientation space [18]. Using a Mathematica© script [19], the obtained Bunge Euler angles were used to generate crystallographic plane traces for the specific grains of interest.

*2.3. Micropillar splitting*

Micropillars were fabricated using an outer to inner pass milling procedure with a focused ion beam (FIB) of both Ga (Zeiss Auriga) and Xe (Thermofischer Helios PFIB) ions. The two ion sources were utilised to mill pillars in order to investigate ion damage effects on the obtained fracture toughness values. For 5 µm diameter $Ga^+$ FIB pillars, annular milling was performed using an acceleration voltage of 30 kV and probe current 2 nA. The current was lowered to 600 pA for fine milling to reduce the impact of ion-damage and re-deposition around the pillar. For the 10 µm pillars, a fine milling current of 2 nA was used. For $Xe^+$ FIB pillars, a course milling current of 15 nA and a fine milling current of 1 nA was employed. Testing was performed using a commercial *in situ* nanoindentation device (ASMEC UNAT-II, Germany) installed in a Zeiss Gemini SEM. A diamond cube corner tip was used for these tests (Synton-MDP, Switzerland), and the measurements were performed in displacement control with 10 nm/s and the tests stopped once failure was initiated.

*2.4. Microcantilever bending*

Rectangular and pentagonal microcantilevers for the bending experiments were milled with 30 kV and 2 nA current using the $Ga^+$-FIB, while a final polishing step with 240 pA was used for each beam to remove any damaged material and redeposition. The dimensions of the rectangular beams were maintained at 15 µm length ($L$), 3 µm width ($W$), and 3 µm breadth ($B$). These dimensions are annotated on a SEM image of a microcantilever in **Fig. 6b**. Beams were milled in two different grains, four on an approximate (111) surface and two on a ~(113) surface orientation, evaluated using EBSD prior to milling. Pentagonal beams were prepared in two different grains with surface orientation (1 1 10) and $(\bar{2}\bar{1}4)$ targeting alignment of the crack plane with low-index planes in the $C$15 crystal structure. Final notching was achieved using a line scan applying 50 pA current for 30 s, for a targeted notch depth ($a$) equating to a ratio of $a/W \approx 0.3$ (for rectangular cantilevers) or $a/2c \approx 0.3$ (for pentagonal cantilevers, where $c$ is the vertical distance between the top surface and the neutral plane). Testing was performed with the ASMEC *in situ* nanoindenter equipped with a diamond wedge indenter (length 10 µm, tip radius ~200 nm, Synton-MDP Switzerland). Tests were performed in displacement control mode with displacement rate of 5 nm/s. The distance between the notch and indenter contact was maintained at $L$ = 12 µm (and $L/W$ ~4) for rectangular beams and $L$ = 11 µm for the pentagonal beams. Both load-time and displacement-time data were smoothed for noise using a Savitzky-Golay filter.

*2.5. Atomistic simulations*

Any atomistic simulations critically depend on the choice of the interaction potential. Especially when studying fracture, care must be taken that the potential correctly reproduces the bonding situation around the highly-stressed crack tip [20]. Initial fracture tests with a modified embedded atom method (M-EAM) potential for the Ca-Al system [21], which was successfully used for simulations of plasticity in $CaAl_2$ [22], showed artificial phase transformations at the crack tip. Therefore, we performed all fracture simulations on a model $NbCr_2$ $C15$ Laves phase, for which an EAM potential was specifically fitted to replicate crack advance [23]. The fracture toughness and lattice trapping range were determined following [24, 25]. The stress intensity factor $K_I$ was controlled by displacing all atoms according to the linear elastic anisotropic solution of a semi-infinite sharp crack under plane strain, mode I loading [26]. The atomically sharp crack thereby is located at the centre of a cylinder of radius $R$ = 22 nm, with the $x$, $y$, and $z$ directions aligned along the crack-propagation, crack plane normal, and crack-front directions, respectively. Atoms with a distance larger than $R$ of 2 nm from the centre are fixed, and periodic boundary conditions are applied along the $z$ (crack front) direction. After minimisation by the FIRE algorithm [27], the load is increased in increments of $\Delta K_I$ = 0.005 MPa·√m and again minimised in an iterative process until the crack advances, which determines the critical load, or atomistic fracture toughness value $K_{IC}$. All molecular statics simulations are performed using the Large-scale Atomic/Molecular Massively Parallel Simulator (LAMMPS) [28] and OVITO [29] was used for analysis and visualisation.

## 3. Results and discussion

*3.1. Microstructural and chemical characterisation*

The microstructure of the as-cast material exhibited significant compositional gradients and a large number of cracks and defects as observed by EPMA and SEM (for a representative SEM image, see Supplementary Information **Fig. S1a**). After annealing at 600°C for 24 hours, the initial defect structure of the material was significantly improved (**Fig. S1b**), and chemically homogeneous equilibrium phase(s) were obtained. The BSE image of **Fig. 1a** shows the annealed microstructure with the matrix phase having typical grain sizes >100 µm and with a small volume fraction of a second phase precipitated along the grain boundaries. EPMA analyses confirmed the matrix phase to be the $CaAl_2$ Laves phase, while the grain boundary phase was found to be $CaAl_4$. EBSD showed the matrix phase to be cubic *C*15 $CaAl_2$ as was confirmed by comparing with data from a crystal structure database [30]. No textural preferences were observed, and grains extended toward the surface of the cylindrical cast piece. Microgeometries for small-scale fracture toughness testing were next prepared on the specimen surface. Positions for nanoindents were selected, and sets of micropillars and microcantilevers were milled in grains of known orientation to control against variations in surface orientations of the polycrystal. The respective results are discussed in the following sections.

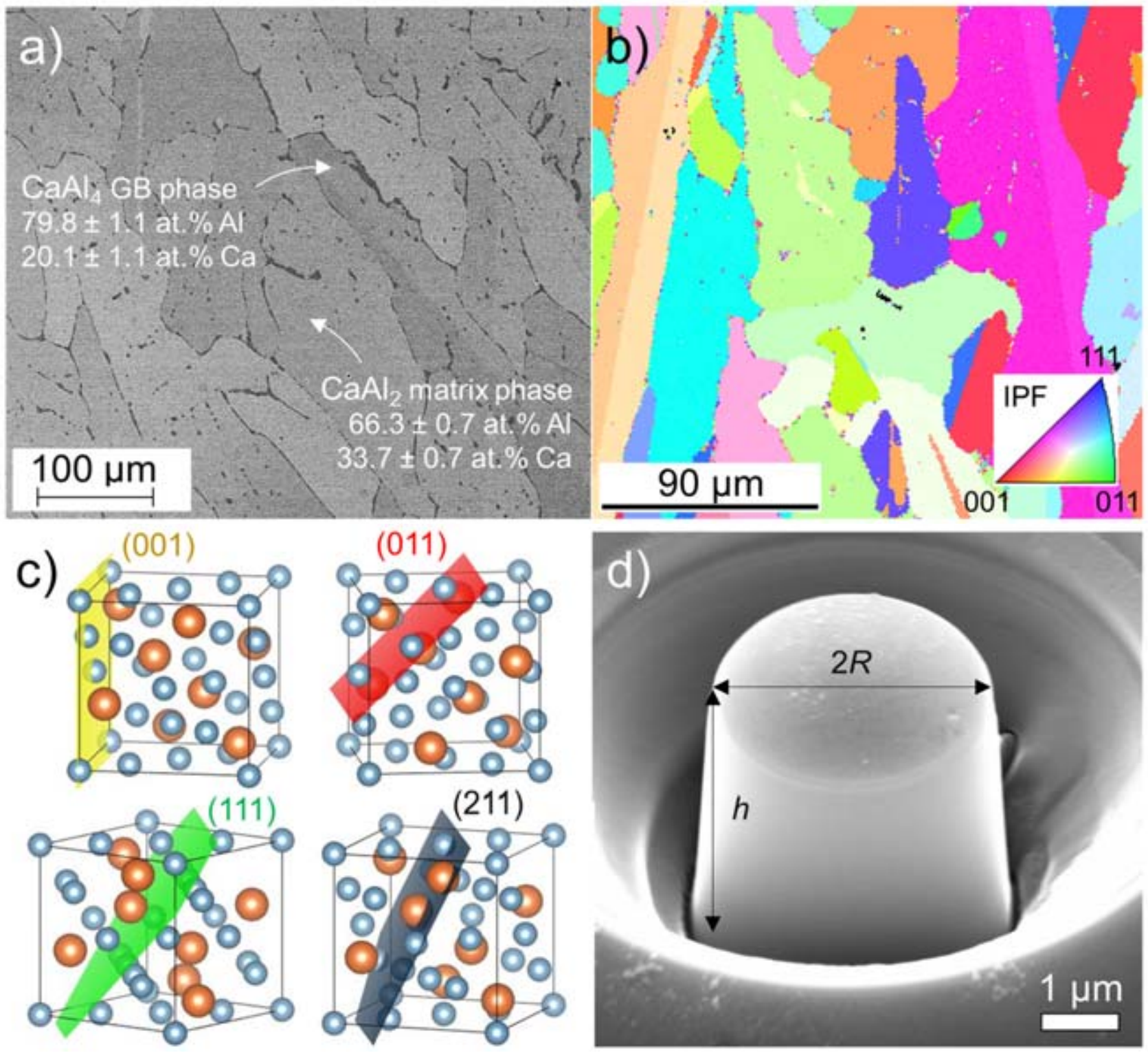

**Figure 1**. Structural and chemical characterisation of the heat-treated *C*15 $CaAl_2$ Laves phase: (*a*) BSE image of the microstructure of the annealed sample, showing small amounts of the grain boundary (GB)

localised $CaAl_4$ skeleton phase. EPMA results highlight the local composition measured for both the matrix and the skeleton phases; (*b*) EBSD image of the cross-section of the annealed cylindrical rod; (*c*) *Fd*-3*m* unit cell of the cubic *C*15 $CaAl_2$ Laves phase (Al *blue*, Ca *orange*) together with a selection of low-index planes; (*d*) SE micrograph of a micropillar prepared for the annealed sample, together with geometries (radius *R* and pillar height *h*) used for toughness analysis.

### 3.2. Nanoindentation

For purposes of materials characterisation, but also for the following fracture toughness analyses, the hardness and elastic modulus is required. From nanoindentation over several grains pre-characterised using EBSD, the average values for hardness and elastic modulus at 500 nm indentation depth were determined from 25 indents to be $H$=5.8 ± 0.1 GPa and $E$=110 ± 2 GPa, respectively. These values were found to be independent of the surface orientation. The collected load-displacement curves for the indents can be seen in **Fig. S2**, and show excellent consistency between measurements. Also shown in **Fig. S2** are exemplary SE images of Berkovich indents into the as-cast and annealed material, where some cracking could be observed in a small sub-set of indents mainly from the corners of the Berkovich geometry. Indentation using a diamond cube corner counterbody was also performed, with the aim of extracting nanoindentation fracture toughness estimates for the annealed *C*15 phase, however significant damage was accumulated around the indent together with uneven and asymmetric crack lengths (**Fig. S3**), inappropriate for toughness evaluation.

### 3.3. Micropillar splitting

To investigate the fracture toughness, a micropillar splitting technique was utilised on pillars with an aspect ratio (height/diameter) maintained at ~1.0. As discussed in the seminal work by Sebastiani *et al.* [31], such geometrical considerations ensure a relaxed residual stress state. The fracture toughness $K_C$ using pillar splitting can simply be determined as [31]:

$$K_C = \gamma \frac{P_c}{R^{1.5}} \tag{1}$$

The critical load $P_c$ was determined from the load-displacement data collected during measurement, and the micropillar radius $R$ determined from SEM imaging of the final polished pillars. A gauge factor $\gamma$ was determined using the ratio of the indentation elastic modulus ($E$) and hardness ($H$) from the Berkovich nanoindentation measurements of Section 3.2. Using linear fitting and evaluation based on the data set of $\gamma$ against $E/H$ reported in Ref. [32] for a cube corner indenter geometry, the gauge factor for the annealed $CaAl_2$ phase was determined to be 0.8 for an $E/H$ ratio of 19. This factor was used in the pillar splitting fracture toughness **Eq. 1**, and was also found to be consistent for the as-cast material. This equation assumes an idealised, isotropic, elastic-perfect plastic material for the cohesive zone finite element simulations used to estimate instability loads from the micropillar radius and fracture toughness, along with no friction between the indenter and top of the pillar, and a plastic zone size small compared to the pillar diameter [31].

As a preliminary step, four 5 μm diameter $Ga^+$-FIB pillars were milled into the as-cast *C*15 Laves phase sample, and the fracture toughness determined to be 1.22 ± 0.12 MPa·√m, independent of surface orientation. Pillars were observed to generally split evenly. Due to the significant number of defects observed in the as-cast material, however, the remainder of the fracture analysis of the *C*15 $CaAl_2$ phase is based on the annealed bulk Laves phase.

### *3.3.1. Examining the effect of crystallographic orientation*

In determining the fracture toughness of the annealed material, two different grain orientations were selected in which micro-pillars with a diameter of 5 μm were cut using a $Ga^+$-FIB. These measurements allowed a comparison with the as-cast material, together with understanding whether there is an orientation dependence on the toughness from pillar splitting in the annealed material. A weak orientation dependence of the nanoindentation hardness has been previously reported for the hexagonal *C*14 $NbCo_2$ Laves phase, while the hardness of cubic *C*15 $NbCo_2$ appeared to be independent of orientation [14]. Load-displacement curves were collected from the splitting measurements (**Fig. 2a**), and *post mortem* SE images for the 5 μm $Ga^+$-FIB pillars are shown in **Fig. 2b,c**. For both tested grains, ideal three-way splitting was observed in the *post mortem* images, with three cracks propagating from a single point underneath the indenter. The uniform cracking in all three directions of the pillars also suggest negligible anisotropy with respect to the indentation direction in pillar splitting, which is predominantly influenced by the complex stress state beneath the cube corner indenter. It is also noted that deviations in indent placement with respect to the pillar centre were within 10% of the pillar's radius $R$, and no influence on the toughness from the indenter placement should therefore result, as has been demonstrated for pillar splitting of Si [33].

Using the gauge factor ($\gamma$) obtained through nanoindentation, the critical load $P_C$ and the pillar radius $R$, the fracture toughness $K_C$ was evaluated applying **Eq. 1**. Average values of $K_C$ = 1.20 ± 0.17 MPa·√m for *grain 1* (four pillars) and 1.23 ± 0.13 MPa·√m for *grain 2* (four pillars) were obtained. These results indicate that orientation does not have an impact on the measured indentation-based fracture toughness tests in the present case, in line with our reported nanoindentation results of Section 3.2. It is also noted that the toughness values agree with the as-cast result of 1.22 ± 0.12 MPa·√m within a single standard deviation, indicating that the heat treatment did not have a notable effect on the intrinsic phase properties, but rather the bulk properties from a reduction of larger defects and minimisation of the $CaAl_4$ skeleton phase.

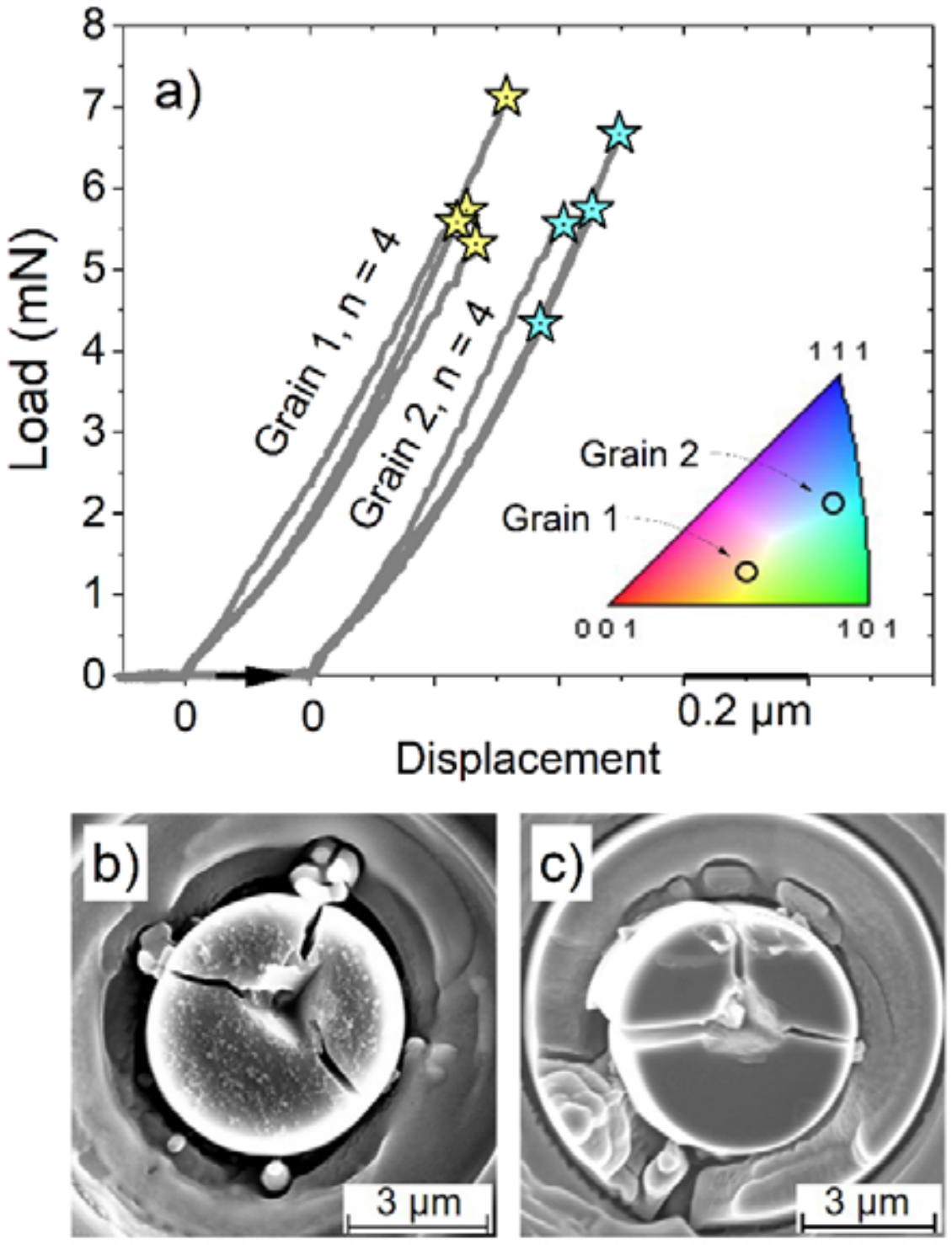

**Figure 2**. Splitting of 5 µm, single-crystal $CaAl_2$ micropillars fabricated in two differently oriented grains using $Ga^+$-FIB. Load displacement curves (*a*) highlight the critical fracture load for grains located as shown in the inverse pole figure (IPF) inset image (*n* indicates total sample number studied). Data for *grain 2* are artificially shifted on the horizontal-axis by 0.2 µm. Representative *post mortem* SE images showing fracture in pillars milled in *grain 1* (*b*) and *grain 2* (*c*).

### *3.3.2. Ion beam and size effects*

A consideration for the toughness evaluation using micropillar splitting are ion beam damage effects from the Ga ions [33]. To understand whether the ion-type used for the FIB milling had an impact on the fracture toughness, 5 µm pillars were milled using $Xe^+$-ions instead of Ga.

These pillars were milled in a single grain with orientation close to that of *grain 1* in **Fig. 2**. **Fig. 3** presents the load displacement curves, together with a *post mortem* image of a 5 µm $Xe^{+}$-FIB pillar. The load-displacement curves show a similar relation as for the $Ga^{+}$-FIB pillars in **Fig. 2a** up to the point of fracture. Using the maximum loads obtained from the load displacement curves, an average fracture toughness of 1.08 ± 0.15 MPa·√m results from the 8 pillars successfully tested. This value lies within the standard deviation of both sets of $Ga^{+}$-FIB pillars, confirming that the choice of ions used for the FIB milling process did not have a significant impact on the fracture toughness values obtained using the pillar splitting technique.

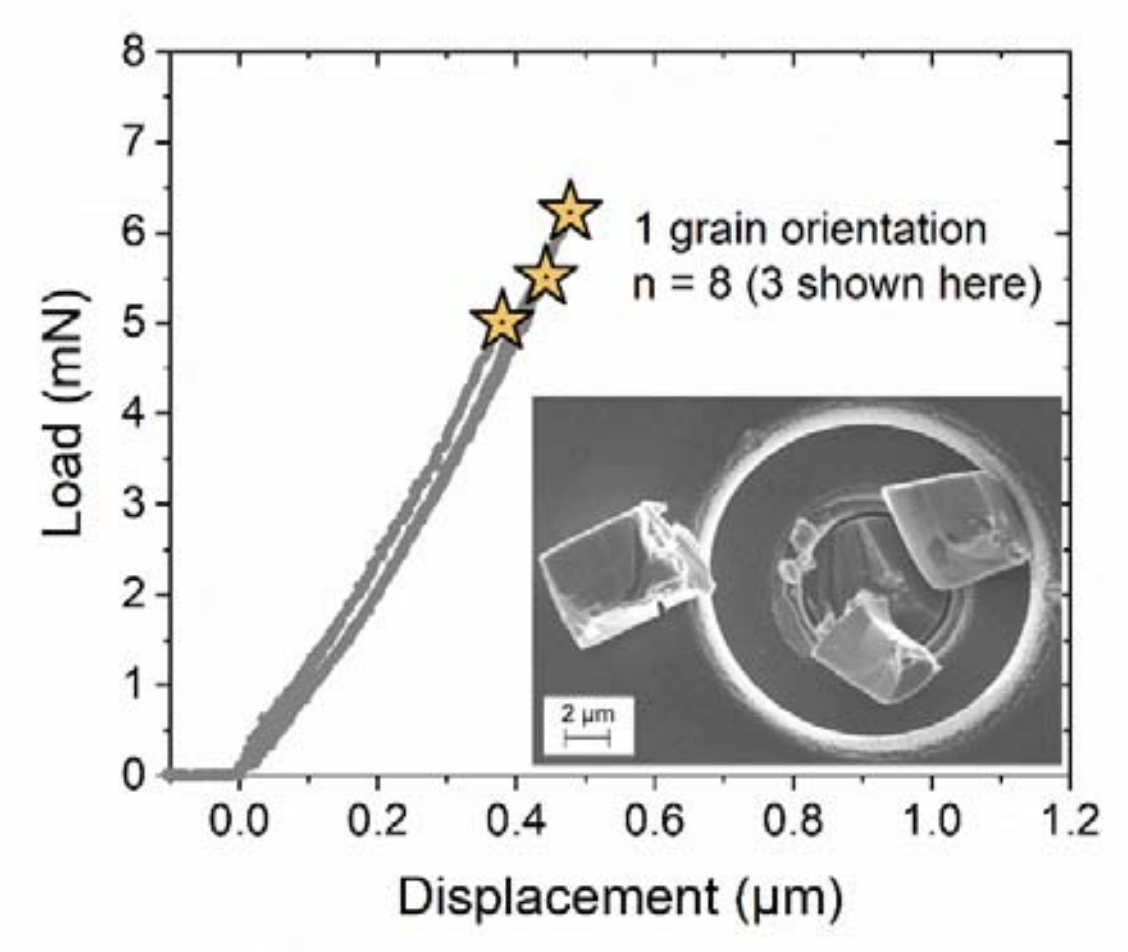

**Figure 3**. Splitting of 5 µm $CaAl_2$ micropillars fabricated using $Xe^{+}$-FIB. Load displacement curves (3 representative curves are shown) highlighting the critical failure load. In the inset a *post mortem* SE image is shown of the fractured pillar, highlighting a separation of the pillar volume into three pieces. A total of 8 successful measurements were made.

Further, pillars of 10 µm diameter were fabricated to rule out size effects on the obtained fracture toughness values, using both $Ga^{+}$- and $Xe^{+}$-FIB sources. **Fig. 4** shows load-displacement curves and *post mortem* images of 10 µm pillars fabricated using both FIB machines. In both cases, an ideal three-way split was seen, and the maximum loads were reasonably constant. The fracture toughness of the 10 µm $Ga^{+}$-FIB pillars was determined as 1.02 ± 0.15 MPa·√m, while the 10 µm $Xe^{+}$-FIB pillars had a toughness of 0.97 ± 0.20 MPa·√m. These values also fall within the standard deviation of the previous results on 5 µm pillars. However, the mean value drops by about 20% for the 10 µm $Ga^{+}$-FIB pillars, and about 10% for the 10 µm $Xe^{+}$-FIB pillars as compared to the 5 µm pillars discussed previously. Such a reduction in toughness with increasing pillar diameter is consistent with observations for Si pillars, where at ca. 10 µm pillar diameter the toughness values stabilised [33]. This has been put down to FIB damage effects for smaller pillars, but interactions of the plastic zone with the pillar outer perimeter could also result in artificial toughening. Considering this, the impact of measurement artefacts on the fracture toughness for the *C*15 $CaAl_2$ Laves phase is negligible as the pillar size increases to 10 µm.

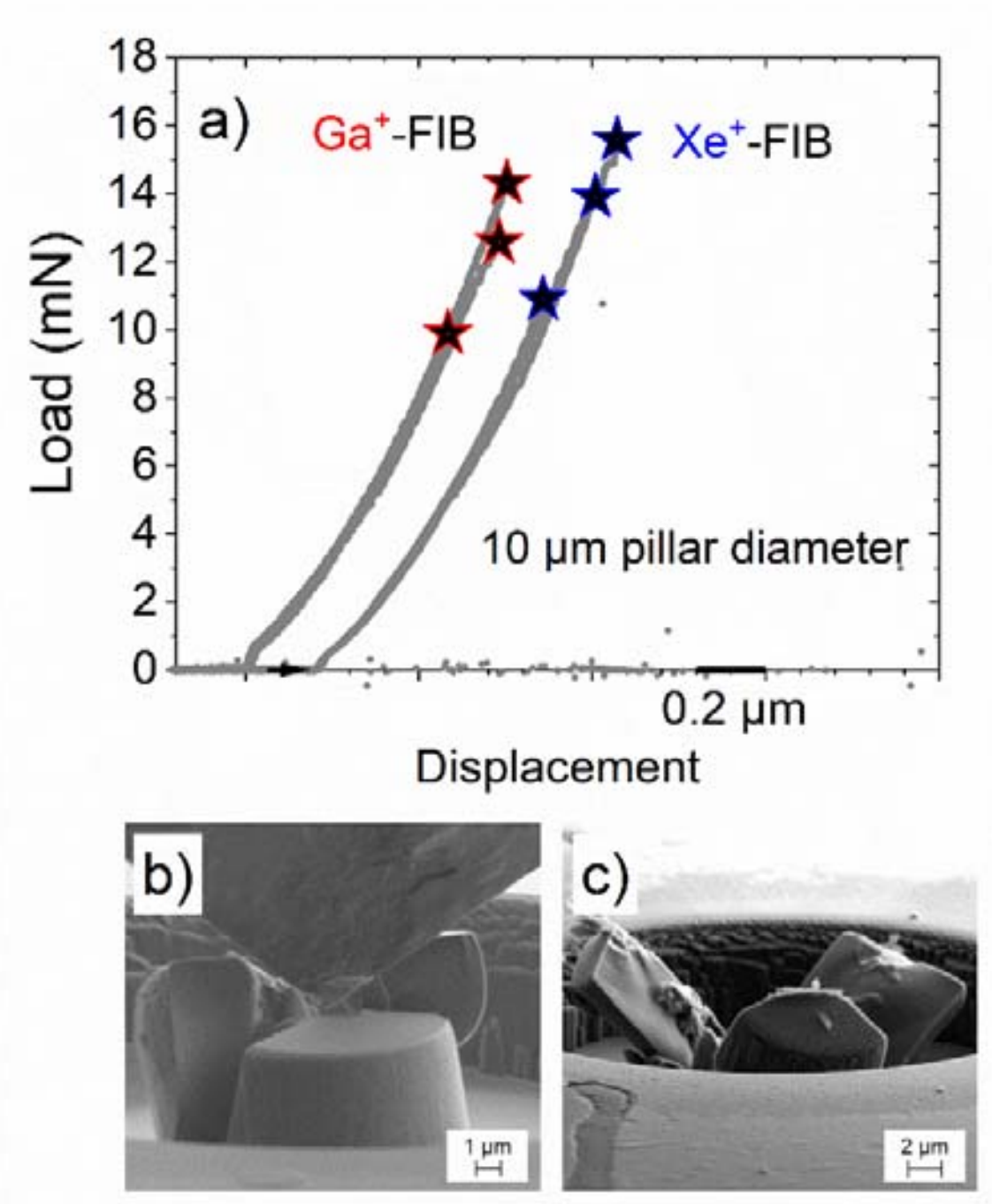

**Figure 4**. Micropillar splitting of 10 µm diameter pillars. Load-displacement curves highlighting the critical fracture load (*a*), and *post mortem* images for pillars milled using $Ga^+$-FIB (*b*) and $Xe^+$-FIB (*c*). Data for the $Xe^+$-FIB milled pillars are artificially shifted on the horizontal axis by 0.2 µm.

In combining the micropillar splitting results, it is clear that only minor variability exists between the cases tested (**Table 1**). It is possible that for the 5 µm pillars, ion-beam damage and plastic zone considerations artificially elevated the toughness measured. However, the results are within a single standard deviation. This indicates that the ion-beam used during FIB preparation, pillar diameter, and micropillar orientation have negligible effects on the evaluated $K_C$ from pillar splitting with a diamond cube corner counterbody. Comparing the results of the as-cast and the annealed sample, it can also be concluded that the annealing does not have a significant impact on the toughness of the *C*15 Laves phase at the microscale. Comparison with indentation based fracture toughness measurements of other *C*15 Laves phases (such as $NbCr_2$, $NbCo_2$, etc., where $K_C \sim 1$ MPa·√m [23, 34, 35]), highlight that the toughness values lie in a similar range as those obtained for the *C*15 $CaAl_2$ Laves phase by the pillar splitting technique (also an indentation-based fracture testing method).

**Table 1**. Overview of toughness results obtained for micropillar splitting of single-crystalline $CaAl_2$. Shown are the mean and single standard deviations for the pillar diameter and toughness determined for the splitting experiments, together with the number of successful tests *n*.

| Treatment | Diameter (µm) | FIB ion type | Toughness (MPa·√m) | *n* |
|---|---|---|---|---|
| As-cast | ~4.57 ± 0.07 | $Ga^+$ | 1.22 ± 0.12 | 3 |
| Annealed | ~4.85 ± 0.27 | $Ga^+$ | 1.22 ± 0.14 | 8 |
| Annealed | ~5.08 ± 0.31 | $Xe^+$ | 1.08 ± 0.15 | 8 |
| Annealed | ~9.03 ± 0.18 | $Ga^+$ | 1.02 ± 0.15 | 3 |
| Annealed | ~10.57 ± 0.10 | $Xe^+$ | 0.97 ± 0.20 | 5 |

### *3.4. Fracture plane analysis from spherical nanoindentation*

Spherical nanoindentation was carried out on three grains with nominal surface planes of (3 2 6), (3 1 10) and (7 4 9) as marked in green, orange and blue diffraction points, respectively, in the inverse pole figure of **Fig. 5a**. Spherical nanoindentation was performed to obtain a symmetrical stress distribution under the indent. The alignment of surface cracks in each of the grains is consistent in all five indents within the same grain. However, these directions differ between the grains. **Fig. 5b-d** show selected ECCI micrographs of indents in three selected grains with a consistent colour code. Cracks are marked with red arrows, while the white lines indicate surface plane traces for a selection of the {011} family. These traces were determined using a slip trace analysis done by an in-house developed Mathematica code. It must be noted that less than 5% of the cracks are aligned with {112} planes (an example is shown in **Fig. S4**), while >95% of the cracks are aligned with {011} planes. The full planes are given in the pole figures of **Fig. 5e-g**, while the dots indicate the diffraction points from the surface of the grain. Evidently, the crack planes match very well with the generated plane traces corresponding to the {011} set, indicating a preferential plane for cleavage fracture. This is an intriguing result, as *ab initio* calculations of the surface energies and critical stress for *C*15 $NbCr_2$ on (100), (110) and (111) cleavage planes show that the (111) plane has the lowest surface energy [36], although the toughness was not calculated. In the results of **Fig. 5**, however, it is found that the preferential cracking on {110} planes during spherical nanoindentation is largely insensitive to crystallographic orientation. Using the stress field of a spherical indenter in isotropic elasticity [37] the opening stresses on different crystallographic planes was calculated. This showed similarly no strong dependence on crystallographic orientation, with the {110} not being the most stressed planes (see **Fig. S5** in the Supplementary Information). These findings are strong indications that {110} planes are the natural cleavage planes of *C*15 Laves phases.

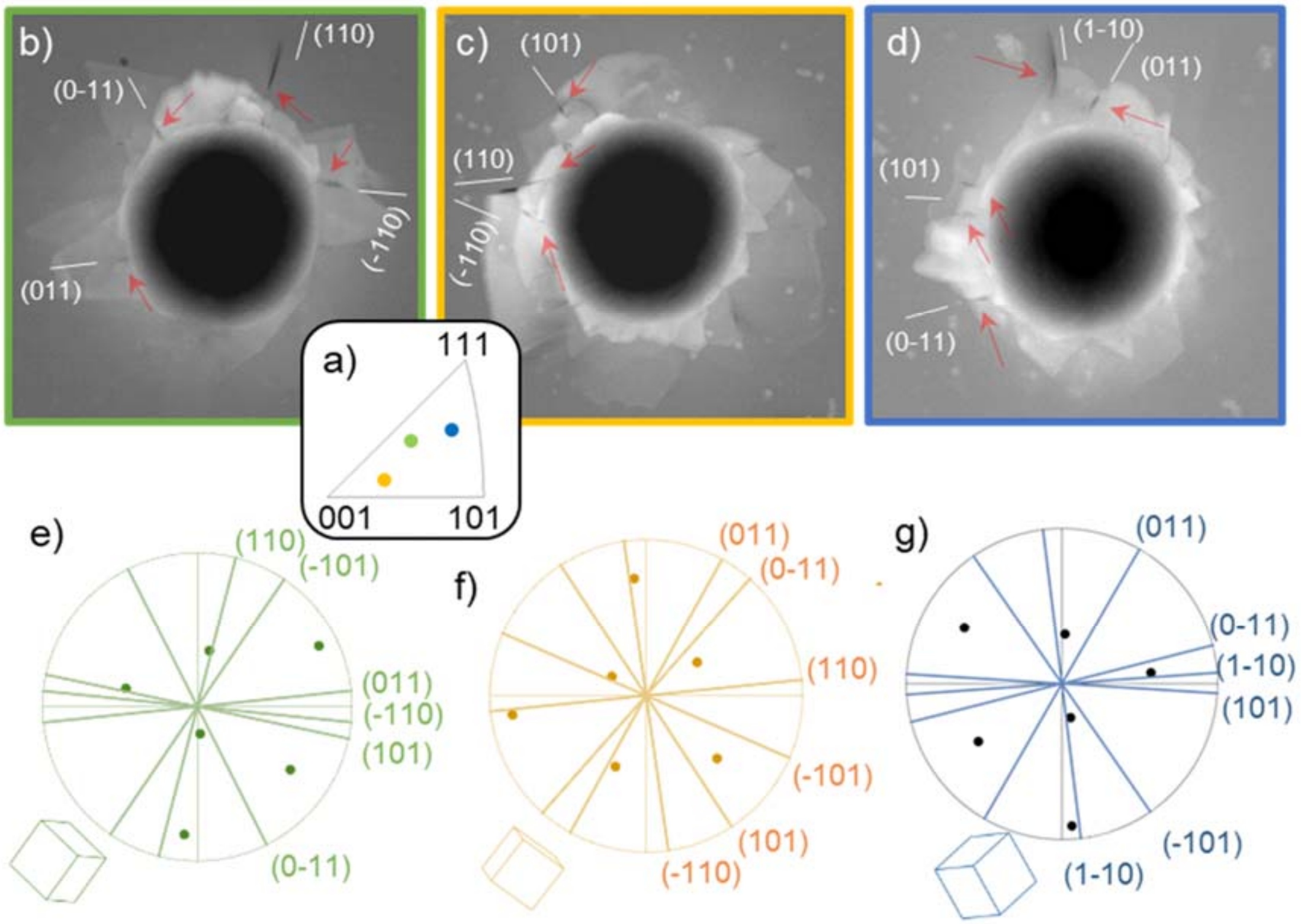

**Figure 5**. Spherical nanoindentation investigation of fracture complimented with ECCI with 0° tilt. Inverse pole figure of 3 selected grains for nanoindentation (*a*). The colour of each diffraction point corresponds to the ECC image and trace analysis with the same colour. Cracks are marked with red arrows, white lines in the SEM images in (*b*), (*c*) and (*d*) imply the trace of the indicated planes as deduced from the trace analysis pole figure in (*e*), (*f*) and (*g*), respectively.

### *3.5. Microcantilever bending experiments*

While a phenomenological relationship between the critical load in pillar splitting and the fracture toughness exists, pillar splitting is not a true fracture mechanics test due to the lack of a pre-existing crack [38]. Experiments on notched microcantilevers allow for a comparison between the fracture toughness quantification from a mode I loaded pre-crack against those from pillar splitting. Moreover, such measurements allow to gain insights into fracture along predefined planes and directions, as well as into possible competition between different cleavage planes. First, a total of six beams were milled in two different grains, four in *grain 1* with a ~{111} surface and two in *grain 2* with a ~{113} surface orientation as indicated in the IPF shown in **Fig. 6a**. The crystallographic direction along the beam length $L$ for both beams was not aligned with any of the low index planes. Further details of the crystallographic orientation along different directions of the cantilevers and corresponding pre-cracks are given later in **Table 3**. **Fig. 6b** shows an SEM image of the notched cantilever beams. As all microcantilevers fail within the linear-elastic region of the deformation curve without showing plasticity (**Fig. 6c**), energy dissipation through plastic deformation is negligible. The fracture toughness for mode I loading can therefore be calculated according to the following equation [15]:

$$K_{Ic} = \frac{P_c L}{B W^{3/2}} f\left(\frac{a}{\mathrm{W}}\right) \qquad (2)$$

where $P_c$ is the critical load where failure occurs, $L$ is the distance between the notch and the point of load application using the wedge indenter, $B$ is the thickness and $W$ is the width of the beam. The dimensions $L$, $B$ and $W$ are depicted schematically in **Fig. 6b**. The dimensionless geometric factor $f(a/W)$ was obtained using an approach for fracture toughness quantification described in detail by Matoy *et al.* [39]. Using **Eq. 2**, the fracture toughness of the cantilevers was 1.88 ± 0.45 MPa·√m for microcantilevers in *grain 1*, and 2.13 ± 0.38 MPa·√m for *grain 2*.

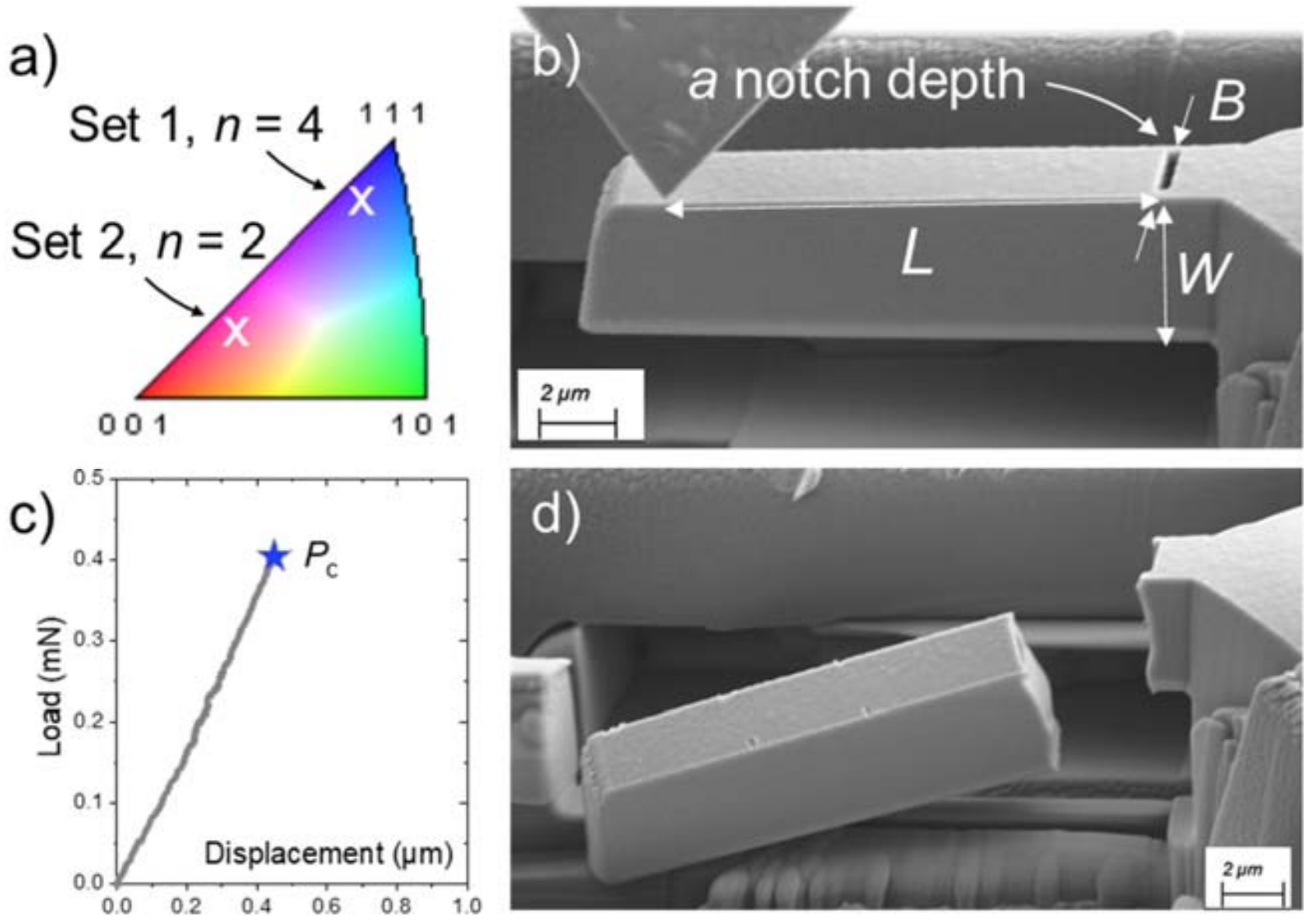

**Figure 6**. Microcantilever fracture experiments. Cantilevers were milled from two grains with out-of-plane orientations indicated in the IPF (*a*), with geometric parameters indicated in the SE micrograph of the cantilever beam (*b*). A representative load-displacement curve for the 5 nm/s displacement-controlled tests is shown in (*c*) highlighting the fracture point. Brittle fracture was observed for all 6 cantilevers tested in both grains based on failure within the linear elastic response (*c*). A *post mortem* SEM image of a fractured cantilever is shown in (*d*).

From **Fig. 6d** it is observed that fracture did not proceed in a purely mode I manner; the fracture plane deviated from the notch plane and does not match any clear single crystallographic plane, as most prominently observed for *grain 1*. From calculations of the misorientation angle between the notch plane and low-index fracture planes (determined by the cross product of the respective plane normals, shown in **Table 2**), the deviation could be analysed in detail. Orienting the *post mortem* SEM micrographs edge-on and tilting at 45° allowed for a direct comparison of the calculated lowest misorientation crystal planes to the deviations of the crack path (**Fig. 7**).

**Table 2**. The four low-index planes with the lowest misorientation angles to the notch plane are listed for both grains studied using microcantilever fracture. Shown is the misorientation angle together with plane.

| | 1st | 2nd | 3rd | 4th |
|---|---|---|---|---|
| *Grain 1* | 8.4° // $(1\bar{1}\bar{2})$ | 27.5° // $(00\bar{1})$ | 27.5° // $(\bar{1}\bar{1}\bar{1})$ | 28.1° // $(\bar{1}0\bar{1})$ |
| *Grain 2* | 14.3° // $(\bar{1}12)$ | 21.3° // $(001)$ | 30.7° // $(011)$ | 33.6° // $(\bar{1}11)$ |

For *grain 1* (**Fig. 7a,b**) the fracture surface is closest to the $(1\bar{1}\bar{2})$ plane, which according to **Table 2** has the least misorientation (8.4°) of the possible low-index planes. For *grain 2* (**Fig. 7c,d**) fracture proceeds approximately in mode I and the fracture surface does not align with any of the lowest misorientation projections. However, here the lowest misorientation is 14.3° with the $(\bar{1}12)$ plane, and it is possible that the crack progresses through a combination of crack planes. The fracture surfaces of **Fig. 7b,d** appear smooth and a detailed analysis of the crack trajectory is therefore not possible.

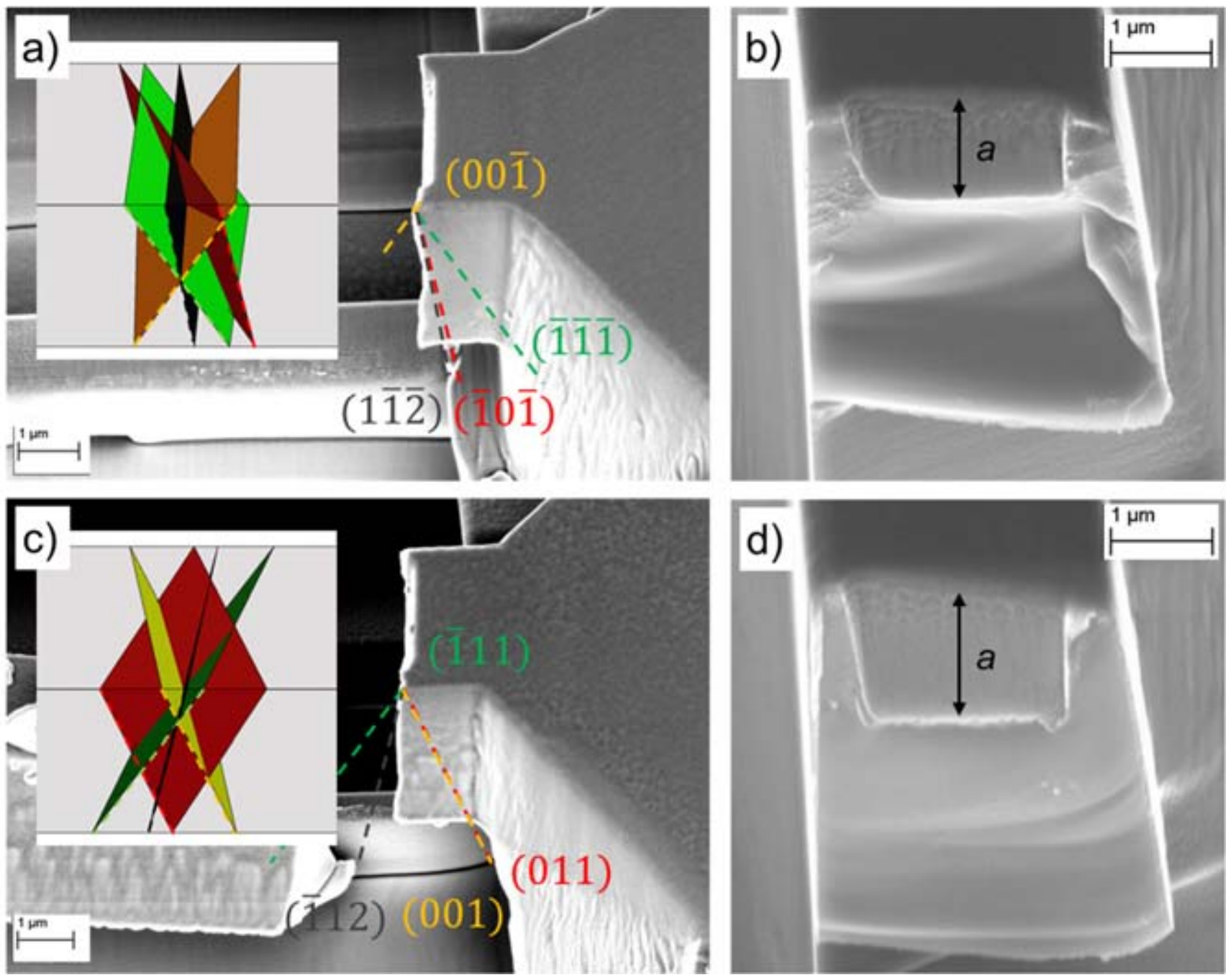

**Figure 7**. Fracture surface analysis for tested microcantilevers in *grain 1* (*a*,*b*) and *grain 2* (*c*,*d*). From SE images of the side-on beams at a tilting angle of 45° (*a*,*c*) potential fracture planes with lowest misorientation angle to the notch plane are shown determined from EBSD analysis (*inset*), while traces of these planes are overlayed onto the micrograph. Tilt-corrected SE micrographs showing the notch depth *a* and fracture surface are additionally provided (*b*,*d*).

Additional fracture toughness tests were performed using pentagonal microbeams away from the sample edge, where the notch plane was aligned with two low-index crystallographic planes, $\{110\}$ and $\{112\}$. The pentagonal beams were milled on two grains having different surface orientation for each of the low-index planes. The cantilevers with notch parallel to the $(\bar{1}10)$ plane were prepared in *grain 3* with surface orientation of (1 1 10), whereas those parallel to the (121) plane were prepared in *grain 4* with surface orientation of $(\bar{2}\bar{1}4)$. **Fig. 8a** shows such a pentagonal cantilever prior to testing inside the SEM and the inset in the figures shows the pentagonal cross-section of the beams. The relevant dimensions are annotated in the figure. **Fig. 8b** shows representative load-displacement curves for the two notch planes, and the IPF in the figure shows the corresponding grain orientations for the two sets of cantilevers. The load displacement curve is linear, indicating that both sets of beams deformed in a linear-elastic manner, similar to the rectangular cantilevers in **Fig. 6**. This is further evident from the fracture surfaces of the two notch systems in **Figs. 8c,d** which show flat surfaces indicating brittle failure. However, no crystallographic facets can be noticed. The fracture toughness of the pentagonal beams was determined using the following equation [15, 40]:

$$K_c = \frac{P_c L c}{I}\sqrt{\pi a} f\left(\frac{a}{2c}\right) \quad (3)$$

Where $P_c$ is the maximum load where failure occurs and is determined from **Fig. 8b**, $L$ is the distance between notch and the loading point, $I$ is the moment of inertia of the cross section of beam, $a$ is the notch depth determined from the fracture surfaces in **Fig. 8c,d**, while the distance $c$ between top surface and neutral plane of the beam is computed according to **Eq. 4** [15, 40, 41].

$$c = \frac{h^2 + hb + c^2}{3(h+b)} \quad (4)$$

The geometric factor $f(a/2c)$ for the pentagonal beam as computed by Chan *et al.* in Ref. [40] is given in **Eq. 5**:

$$f\left(\frac{a}{2c}\right) = 3.710\left(\frac{a}{2c}\right)^3 - 0.63\left(\frac{a}{2c}\right)^2 + 0.242\left(\frac{a}{2c}\right) + 0.974 \quad (5)$$

The value of $a/2c$ was ~0.3 and similar to the $a/W$ ratio of the rectangular beams. Using **Eq. 3**, a fracture toughness $K_{\mathrm{Ic},\{110\}} = 1.80$ and $1.89\ \mathrm{MPa}\cdot\sqrt{\mathrm{m}}$ was obtained for the two beams with $\{110\}$ notch plane, and $K_{\mathrm{Ic},\{112\}} = 1.87$ and $2.17\ \mathrm{MPa}\cdot\sqrt{\mathrm{m}}$ was obtained for the two beams with $\{112\}$ notch plane, both of which are similar to the toughness obtained earlier from rectangular cantilevers. It is worth mentioning that the cantilever geometry, *i.e.* rectangular or pentagonal cross-section, does not influence the linear-elastic fracture toughness values obtained, in alignment with Luo *et al.* for *C*15 $NbCo_2$ [15]. The microcantilever results suggest that the fracture toughness of the *C*15 $CaAl_2$ Laves phase does not vary significantly with notch plane orientation and is of the order of $K_{Ic}$ = 2.0 ± 0.5 MPa·√m. It needs to be emphasized, however, that due to experimental constraints the study of specific crack systems characterised by both low-index crack plane *and* the crack front directions was not possible. It is well known that the cleavage process is anisotropic with respect to propagation direction within a specific cleavage plane. The influence of the propagation direction on the fracture toughness can have a similar magnitude to that of the crack plane [42, 43]. It seems unlikely but cannot be excluded, that well-defined crack systems with lower fracture toughness do exist in the *C*15 $CaAl_2$ Laves phase that have not been tested here.

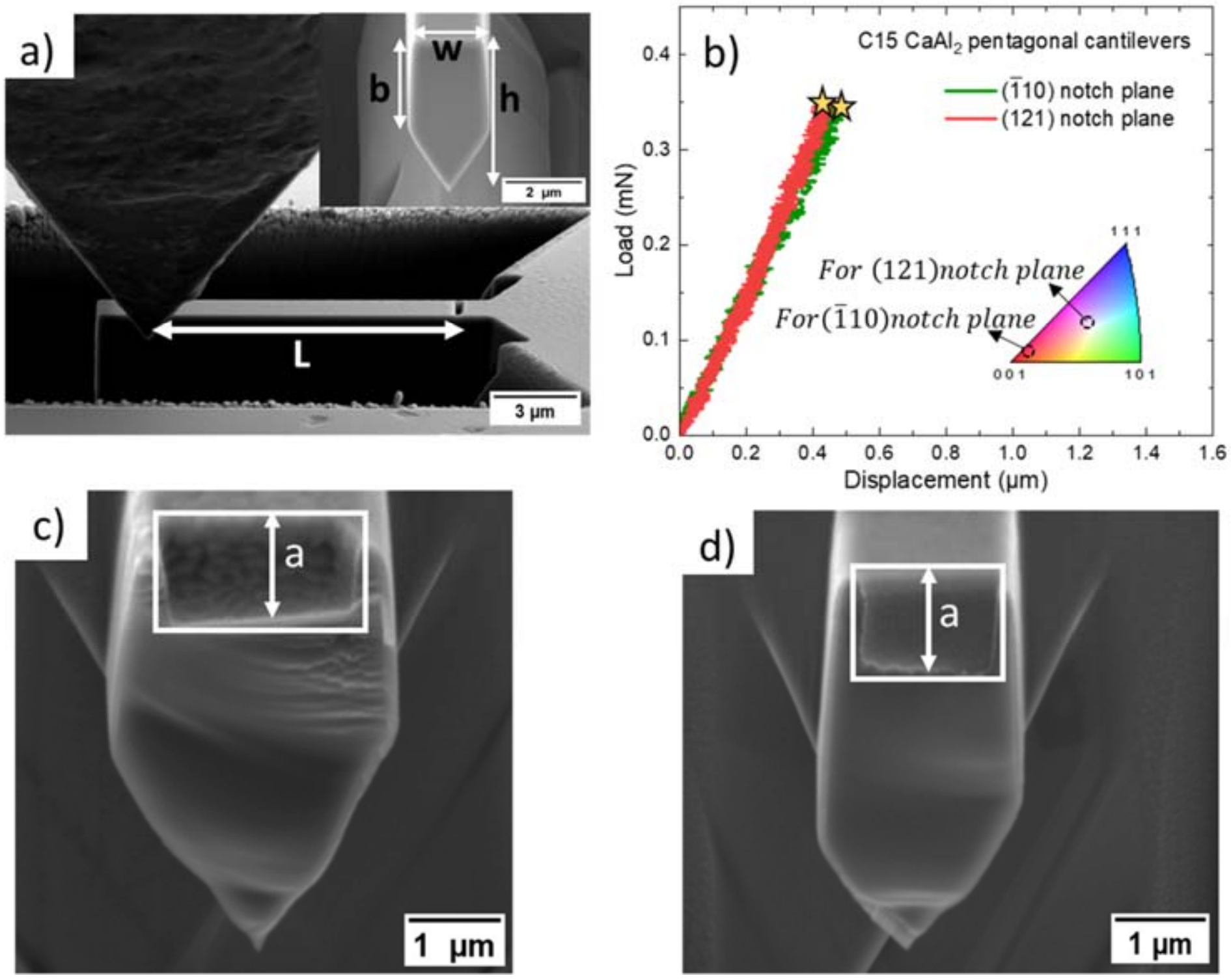

**Figure 8**. Fracture toughness testing of low-index planes using notched pentagonal microcantilevers. (a) Representative pentagonal cantilever inside SEM before testing; inset shows the pentagonal cross-section of the beam with the relevant dimensions annotated in the figure, (b) Representative load-displacement curves of the two sets of pentagonal beams with the grain orientations for each case shown in the IPF (inset), and corresponding tilt corrected SE images of the cantilever fracture surfaces (taken

at 45° tilt angle) for (c) $(\bar{1}10)$ and (d) $(121)$ notch planes showing flat surface due to brittle failure. The notch length *a* is highlighted in both cases.

**Table 3**. Summary of microcantilever fracture toughness values for *C*15 $CaAl_2$. Refer to **Fig. S6** in Supplemental Information for definition of notch plane and notch front. * indicates cross-product derivation.

| Cantilever geometry | Notch plane | Notch front | Direction perpendicular to notch front | Fracture toughness MPa·√m | *n* |
|---|---|---|---|---|---|
| Rectangular, *Grain 1* | $(\overline{50}\,\overline{40}\,\overline{123})$* | $[4\,\bar{5}\,0]$ | $[15\;\;12\,\overline{10}]$ | 1.88 ± 0.45 | 4 |
| Rectangular, *Grain 2* | $(\overline{205}\;257\;844)$* | $[\overline{41}\,\overline{13}\,\bar{6}]$ | $[\bar{5}\;19\;\bar{7}]$ | 2.40; 1.86 | 2 |
| Pentagonal, *Grain 3* | $(\bar{1}10)$ | $[10\;10\;\bar{2}]$* | $[1\;1\;10]$ | 1.80; 1.89 | 2 |
| Pentagonal, *Grain 4* | $(1\;2\;1)$ | $[3\;\bar{2}\;1]$* | $[\;\bar{2}\;\bar{1}\;4]$ | 1.87; 2.17 | 2 |

*3.6. Insights from atomistic simulations*

In the case of negligible plastic deformation, the theoretical fracture toughness according to Griffith $K_{IG}$ is related to the critical energy release rate $G_C$, *i.e.* the change in elastic strain energy per unit area of crack advance, by:

$$K_{IG} = \sqrt{E^* G_C} \qquad (6)$$

Where $E^*$ denotes an appropriate elastic modulus [38]. For brittle fracture $G_C = \gamma_1 + \gamma_2$, with $\gamma_{1,2}$ being the energies of the two surfaces created by crack advance. Due to the different possible terminations of the {111} planes, these can be different in *C*15 Laves phases (see **Fig. S7** in the Supplementary Information). With the material properties from **Table S1** $K_{IG}$ can be calculated for different crack systems, including different terminations, for our *C*15 $NbCr_2$ model system (**Table 4**). For $CaAl_2$, $K_{IG}$ as determined from the DFT calculations in Ref. [36] are presented in **Table S2**. It can be clearly seen that {111} planes with the termination A (table 4, Fig. S7) should have, in thermodynamic equilibrium, the lowest fracture toughness.

The critical fracture toughness $K_{IC}$ values obtained by direct *K*-controlled atomistic simulations, are also shown in **Table 4**. All crack systems showed brittle cleavage without any indication of dislocation activity. Examples of crack tip configurations before and after crack advance (*i.e.* bond breaking) are shown in **Fig. S8**. In all cases, the atomistically determined $K_{IC}$ are larger than the corresponding $K_{IG}$. In addition, $K_{IC}$ is now actually the lowest for {110}

planes – in accordance with experimental observation of cracks along {110} planes instead of {111} plane during spherical nanoindentation.

**Table 4**. Theoretical, thermodynamic fracture toughness according to Griffith, $K_{\mathrm{IG}}$, and fracture toughness values, $K_{\mathrm{IC}}$, as determined from quasistatic atomistic simulations for various crack systems in the $C$15 $NbCr_2$ Laves phase (in units of $MPa \cdot \sqrt{m}$). The lowest $K$ values are reported in bold. The definition of the terminations A and B are shown in Fig. S7.

| Crack-plane | Crack-front | Termination | $K_{\mathrm{IG}}$ | $K_{\mathrm{IC}}$ |
|---|---|---|---|---|
| (111) | $[\bar{1}10]$ | A | **1.004** | 1.210 |
| (111) | $[\bar{1}10]$ | B | 1.066 | 1.230 |
| (111) | $[11\bar{2}]$ | A | **1.004** | 1.215 |
| (111) | $[11\bar{2}]$ | B | 1.066 | 1.22 |
| $(\bar{1}10)$ | $[\bar{1}\bar{1}2]$ | - | 1.017 | **1.095** |
| $(\bar{1}10)$ | [111] | - | 1.015 | **1.065** |
| $(\bar{1}10)$ | [001] | - | 1.02 | **1.04** |
| $(\bar{1}\bar{1}2)$ | [111] | - | 1.017 | 1.130 |
| $(11\bar{2})$ | $[\bar{1}10]$ | - | 1.019 | 1.160 |

The well-known observations that atomistically determined $K_{Ic}$ are, even for perfectly brittle fracture, larger than $K_{\mathrm{IG}}$ are due to the fact that the latter is based on a thermodynamic treatment of fracture in terms of continuum properties, and neglects the underlying atomic nature of the material [42, 43]. This effect is called lattice trapping and is determined by the forces required to break the bonds directly at the crack tip, *i.e.* by kinetics. An atomically sharp crack tip can remain stable until a load $K^{+}$ larger than $K_{\mathrm{IG}}$ is reached. This load is called the upper trapping limit. Similarly, a lower trapping limit $K^{-} < K_{\mathrm{IG}}$ exists, above which the crack will not heal [42]. The trapping range is defined as $\Delta K = K^{+}/K^{-} - 1$. The lattice trapping range can be quantified from the bonding distance-$K$ plots where bonding distance is the distance between the bonds at the crack tip where the initial breaking occurs and $K$ is the applied $K$ field. An example of this plot is shown in **Fig. S8a** for two crack systems. As can be seen in **Figs. S8a** and **S9a**, the lattice trapping range is much larger for breaking the Nb-Cr crack tip bond for the (111)[11-2] and (112)[-110] crack systems than for the Cr-Cr crack tip bond in the (-110)[111] system ($\Delta K^{(110)} = 0.04 < \Delta K^{(111)} = 0.19$). As can be seen from **Fig. S8b** crack propagation in the (-110)[111] system takes place by breaking only one bond. In contrast, for the (111)[11-2] system the load at the crack tip is distributed amongst many bonds as shown by their relaxation after unloading by crack propagation. Similar behaviour can be observed for the (112)[-110] system as shown in **Fig. S9b**. This shows that the lattice trapping is intimately connected to the local crystal structure at the crack tip, and our results therefore are most likely valid for other *C*15 structures. Lattice trapping was already shown to be the underlying reason for the preference of (100) cleavage over cleavage on the lower-energy (110)-planes in tungsten [43] and to influence the crack propagation direction in Si [42]. Lattice trapping is therefore also most probably the cause for the experimentally observed preference of the {110} fracture plane over the low-energy {111} cleavage plane in *C*15 $CaAl_2$.

## 4. Conclusions

In this study, a detailed micromechanical analysis of fracture in bulk cast *C*15 $CaAl_2$ Laves phase was performed. Annealing of the sample removed bulk-scale defects and led to improved handling of the sample for metallographic preparation and *in situ* nanomechanical testing. The indentation hardness and elastic modulus for the *C*15 phase was determined to be $H = 5.8 \pm 0.1$ GPa and $E = 110 \pm 2$ GPa, respectively.

Using micropillar splitting, values of ~1 MPa·√m were determined for the fracture toughness of pillars of 10 μm diameter, while effects of ion-beam damage, pillar diameter and crystal surface orientation were deemed negligible. For spherical indentation on three grain orientations, the {110} family of crystallographic planes consistently showed preferential fracture, and only a small number of cracks were observed on {112} planes. Fracture experiments using microcantilevers were inconclusive concerning quantitative differences in fracture toughness and fracture plane anisotropy. No crystallographic facets could be identified on the fracture surfaces. While a possible deviation of the crack plane towards a close-lying {112} plane was observed in a rectangular microcantilever, the fracture toughness values of cantilevers with {112} and {110} notch planes were indistinguishable. However, the crack front directions in these fracture tests could not be chosen to be parallel to special, low-index directions. Atomistic simulations showed clearly that {110} had the lowest lattice trapping, suggesting that the preference for cleavage along {110} planes observed during spherical nanoindentation is due to kinetic rather than energetic reasons (which would predict {111} cleavage planes).

The microcantilever results showed that the fracture toughness of the *C*15 $CaAl_2$ Laves phase does not vary significantly with notch plane orientation and is of the order of $K_{IC} = 2.0 \pm 0.5$ MPa·√m. This value clearly deviates from the fracture toughness of $K_C = 1.0 \pm 0.2$ MPa·√m obtained from indentation-based pillar splitting (10 μm diameter, $Xe^+$-FIB) through **Eq. (1)**.

Similar deviations between toughness values from indentation based testing and microcantilever testing were reported for *C*15 $NbCo_2$ Laves phase [15, 34, 35] and for tungsten carbide particles [44]. In contrast, the $K_C$ obtained from microcantilever bending and indentation based pillar splitting for single crystal Si(100) were in good agreement [45]. Such differences can be attributed to the inherently different assumptions of the two testing methods. In the indentation-based pillar splitting technique it is reasoned that at the point of fracture instability, the cleavage planes are directly activated resulting from the complex 3D stress distributions present under the sharp cube corner indenter tip. Furthermore, it is speculated that few dislocations formed below the indenter tip during indentation, in the particular pile-ups needed to create the crack nuclei can lead to anti-shielding (local stress enhancement) effect, meaning that the local stress intensity at the crack tip is much higher than from the far-field indenter loading. Such effects are not accounted for in elasto-plastically isotropic models such as those used to derive **Eq. 1.** Furthermore, the derivation of **Eq. 1** in Ref. [31] does not consider the well-known direction dependence of fracture toughness [44,45]. While the semi-circular cracks in indentation-based techniques sample all propagation directions that lie in the fracture plane, cracks in microcantilever-setups are restricted by the notch direction. I.e. a crack on the same plane could start propagating in a low-toughness direction and provide crack front kinks to the remaining semi-circular crack front at a lower stress intensity factor in indentation-based techniques as a crack with a given, arbitrary high-toughness crack front orientation in the microcantilever.

In conclusion, our results reveal that while the *C*15 $CaAl_2$ Laves phase shows a low fracture toughness value around $K_{IC} = 2.0 \pm 0.5$ MPa·√m, for arbitrary crack systems, kinetic effects like direction- and plane dependent lattice trapping play a role for this complex crystal structure and lead to a preference of {110} cleavage planes.

**Acknowledgements**

The authors gratefully acknowledge the financial support of the Deutsche Forschungsgemeinschaft (DFG) within projects B06 and A02 of the Collaborative Research Centre (SFB) 1394 "Structural and Chemical Atomic Complexity - from defect phase diagrams to material properties" – project number 409476157. Funding by the European Research Council (ERC) under the EU's Horizon 2020 Research and Innovation Program is gratefully acknowledged by AG and EB (ERC Consolidator Grant, microKIc, Grant no. 725483). The *C*15 intermetallic was synthesised within Project S of SFB 1394, and Hauke Springer and Michael Kulse are acknowledged for preparation and heat treatment of the sample, along with Leandro Tanure from the IBF at the RWTH-Aachen University. We would like to further acknowledge Sandra Korte-Kerzel and Martina Freund from the Institute for Physical Metallurgy and Materials Physics at the RWTH-Aachen University for fruitful discussion. Martin Palm and Irina Wossack from MPIE are acknowledged for support during EPMA measurement and analysis, while Christoph Kirchlechner is finally acknowledged for development of the Mathematica script used for crack plane analysis.

## Conflict of interests or competing interests

The authors declare that no conflict of interest or competing interests exist.

## Data and code availability

Underlying data for this publication is available at DOI:10.5281/zenodo.10407253.

## Supplementary information

The Supplementary Information document linked to this manuscript contains SEM images of as-cast and annealed *C*15 phase microstructures, nanoindentation curves and SEM images of residual nanoindentation imprints. The Supplementary Information also provides additional details of material constants for calculating the theoretical fracture toughness, examples of crack tip configurations and lattice trapping effect obtained from the atomistic simulations of model *C*15 $NbCr_2$ system.

## Ethical approval

Not applicable.

**Supplementary Information for**

# Fracture of the *C*15 $CaAl_2$ Laves phase at small length-scales

James P. Best[*,†], Anwesha Kanjilal[†], Alireza Ghafarollahi, Uzair Rehman, Chunhua Tian[a)], Hanna Bishara[b)], M. Kamran Bhat, Leon Christiansen, Erik Bitzek, Frank Stein, Gerhard Dehm[*]

[1] *Max-Planck-Institut für Eisenforschung GmbH, Max-Planck-Str. 1, D-40237 Düsseldorf, Germany*

[a)] *Present address: Laboratory for Mechanics of Materials and Nanostructures, Empa – Swiss Federal Institute for Materials Science and Technology, CH-3603 Thun, Switzerland*

[b)] *Present address: Department of Material Science and Engineering, Tel Aviv University, Ramat Aviv 6997801, Tel Aviv, Israel*

[†] *These authors contributed equally to this work*

* *Corresponding authors: j.best@mpie.de and g.dehm@mpie.de*

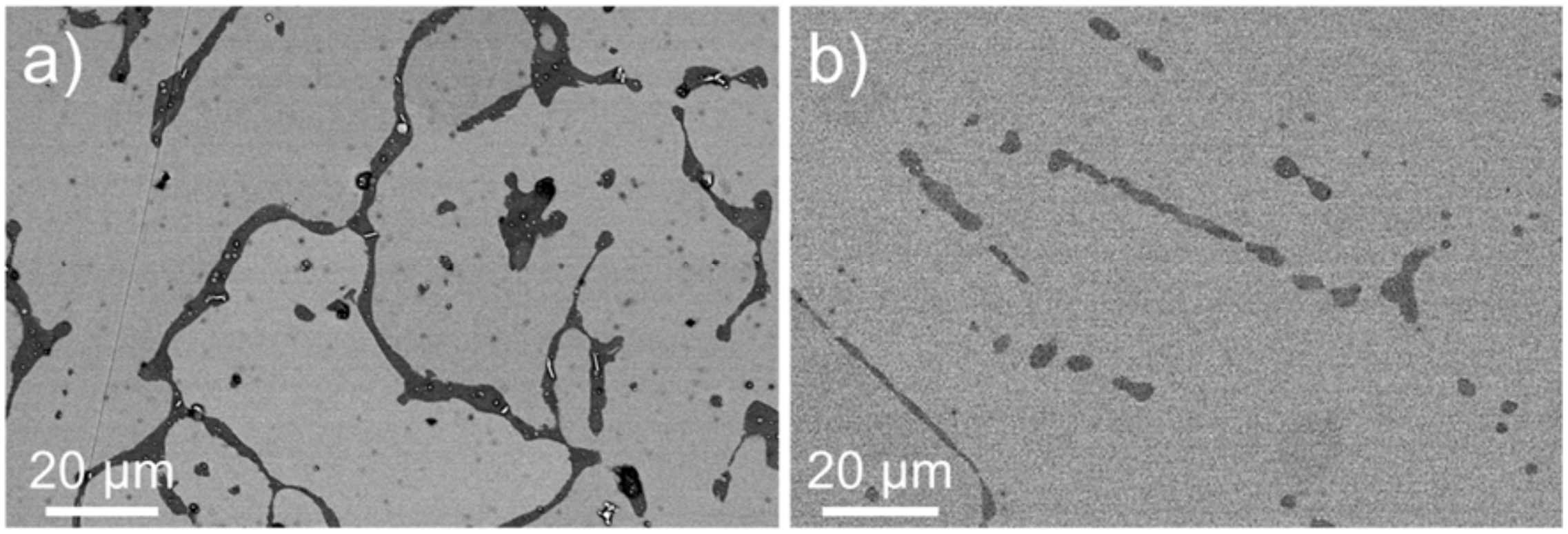

**Figure S1**. BSE microstructure images of as-cast (a) and annealed (b) $CaAl_2$, showing significant reduction of the grain boundary localised $CaAl_4$ skeleton phase and reduction of defects after annealing.

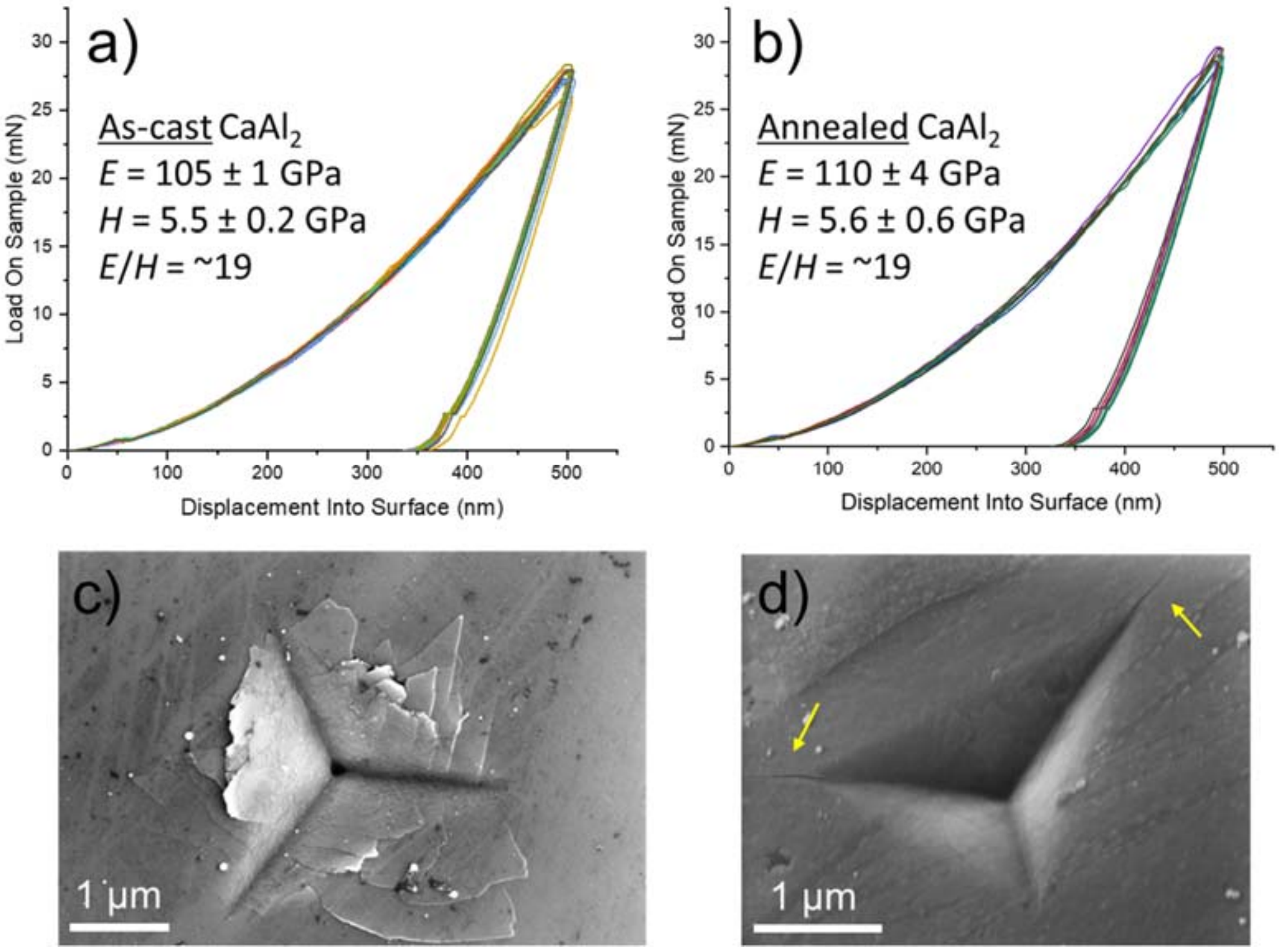

**Figure S2**. Berkovich nanoindentation analysis of both as-cast (*a*,*c*) and annealed (*b*,*d*) *C*15 phase. These *E* and *H* values for the annealed sample were found to be independent of the surface orientation. The collected load-displacement curves for the indents can be seen (*a*,*b*), and show excellent consistency between measurements. Also shown are exemplary SE images of Berkovich indents into the as-cast (*c*) and annealed (*d*) material, where some cracking could be observed in a small sub-set of indents primarily from the corners of the Berkovich geometry (yellow arrows in *d*).

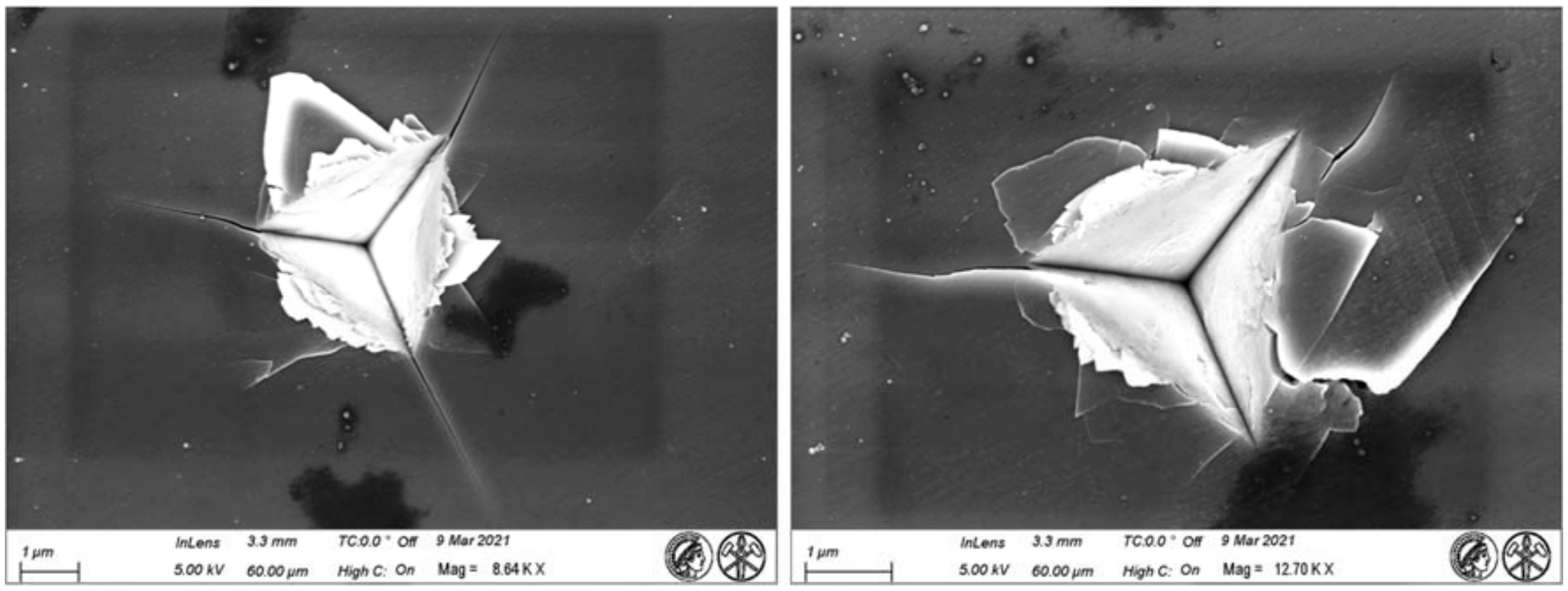

**Figure S3**. SE micrographs of *post mortem* cube corner indentation to 2 µm depth on the annealed C15 phase. A complex damage profile is accumulated around the indent.

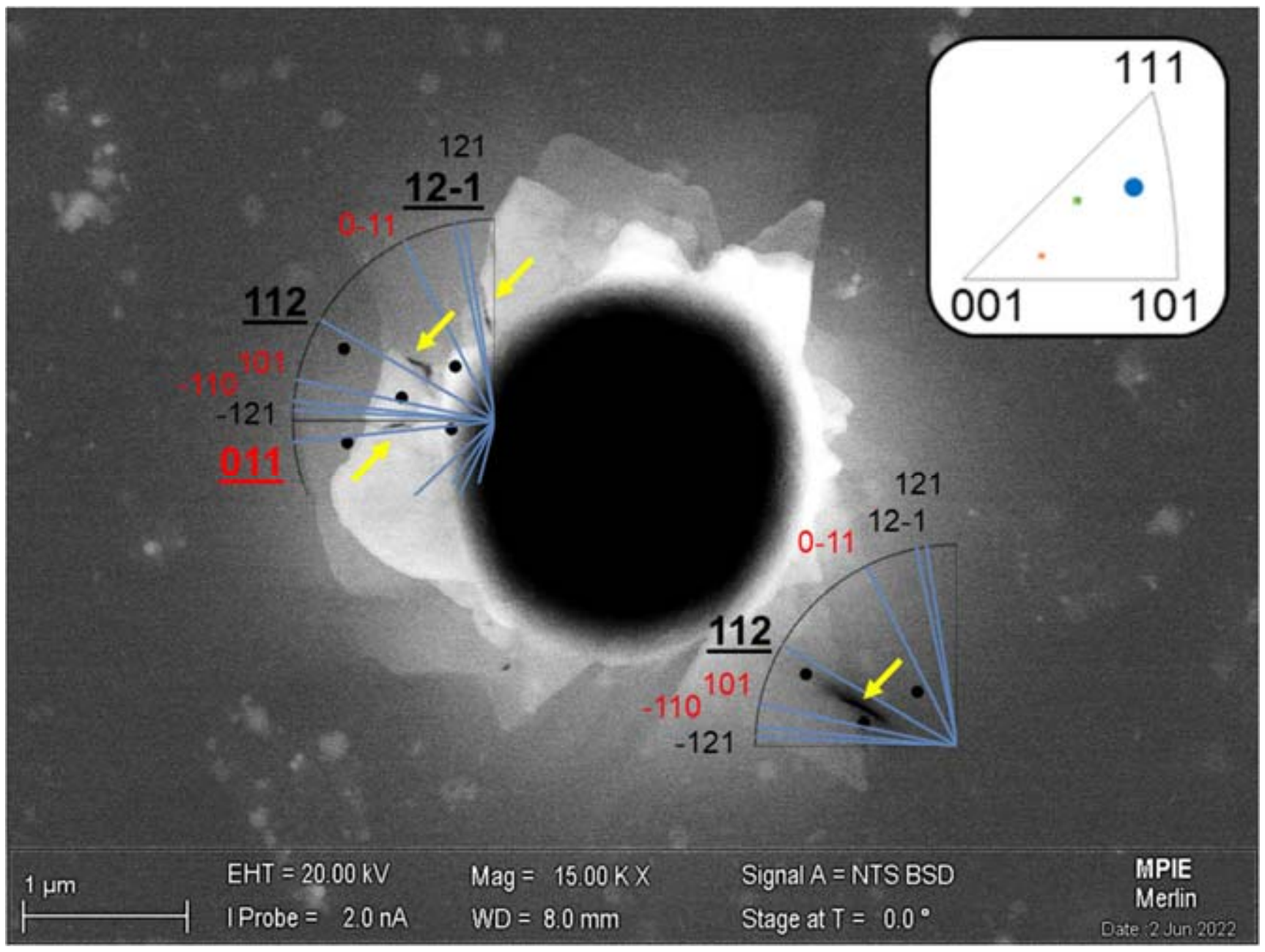

**Figure S4**. Example of indent into *grain 3* (see **Fig. 6d,g**) where the ECCI micrograph shows cracks presenting on the {112} family highlighted with yellow arrows. Presumed cracks on {112} also seen for a single indent each on *grain 1* and *grain 2* (**Fig. 6b,e** and **Fig. 6c,f**, respectively), however in total only 3/15 indents showed evidence of cracks on {112}, and {110} cracks dominated.

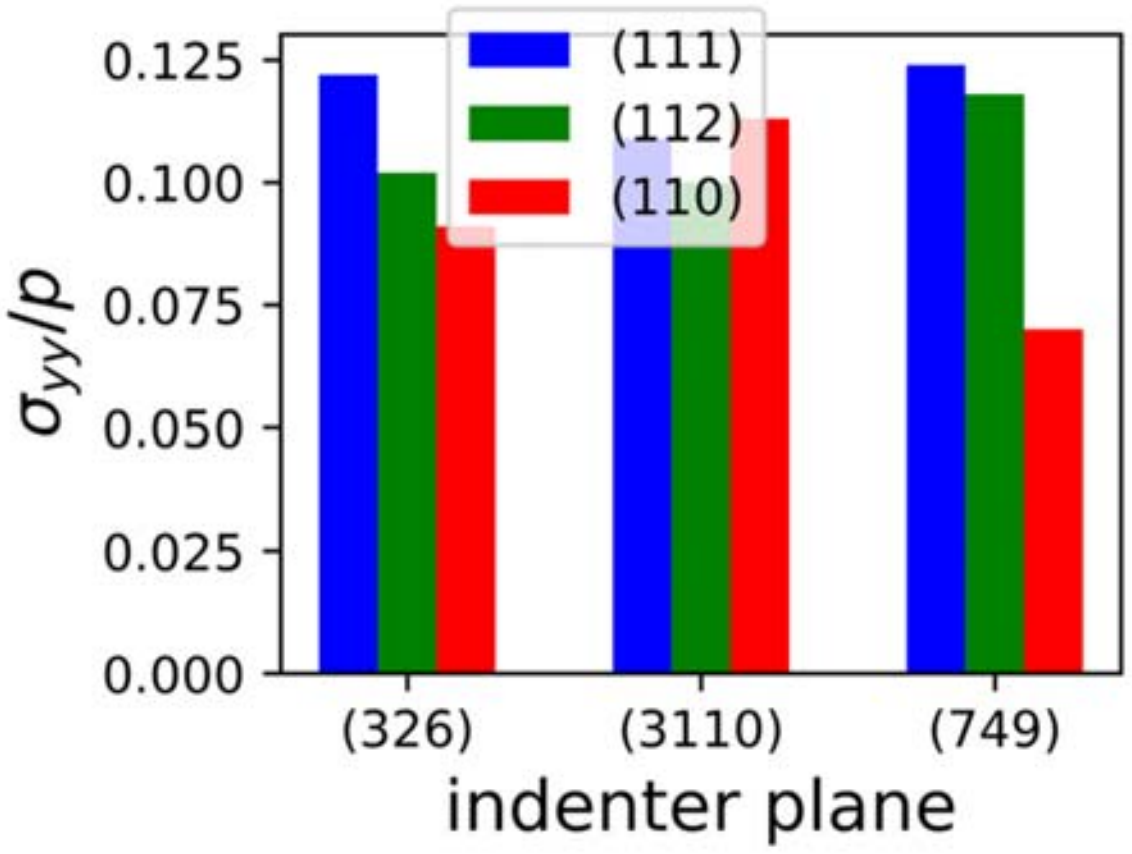

**Figure S5**. Maximum opening stress $\sigma_{yy}$ on possible cleavage planes for the three tested grain surface orientations, i.e., (3 2 6), (3 1 10) and (7 4 9), caused by spherical indent with pressure $p$ according to the isotropic, linear elastic solution in Ref. [37].

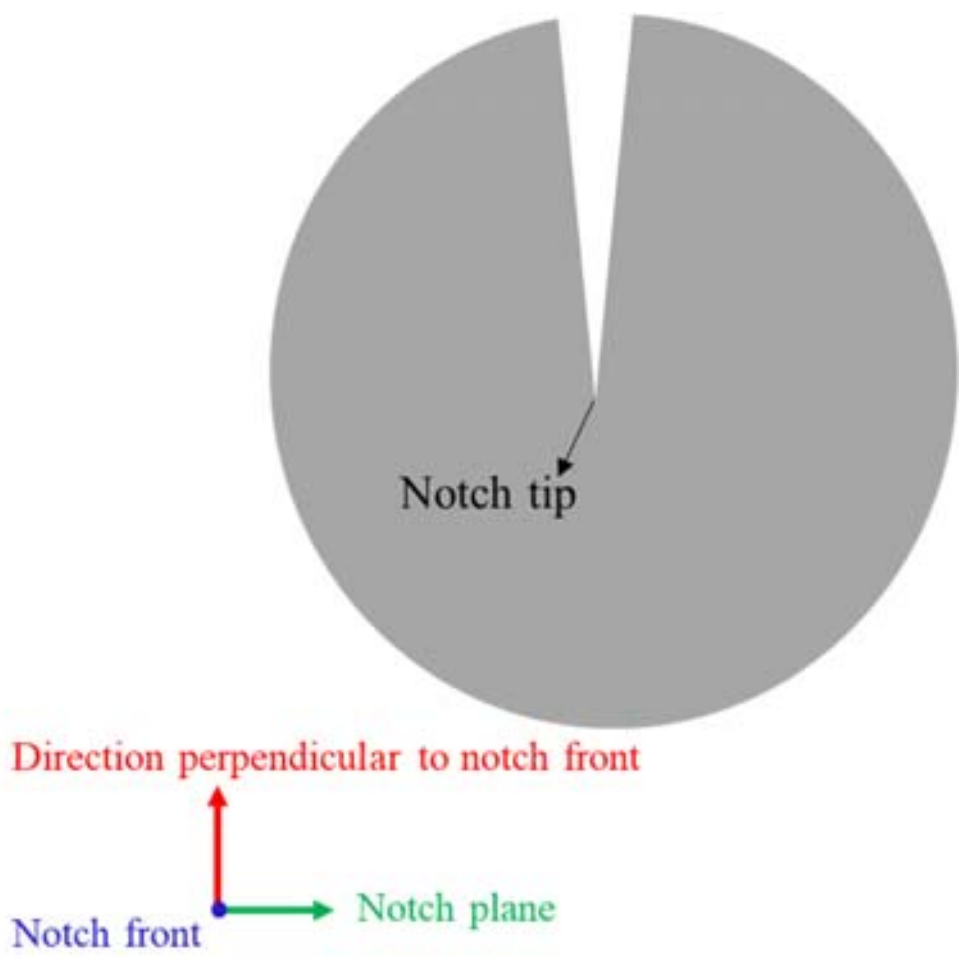

**Figure S6**. Schematic illustration of a pre-crack showing the directions of notch plane and notch front.

**Table S1**. Elastic constants, $C_{ij}$ (in units of GPa) and the critical energy release rate along different atomic planes, $G_c$ (in units if $J/m^2$) as computed by the EAM potential for our $C15$ $NbCr_2$ model system.

| Material | $C_{11}$ | $C_{12}$ | $C_{44}$ | $G_c$ {111}A | $G_c$ {111}B | $G_c$ {112} | $G_c$ {110} |
|---|---|---|---|---|---|---|---|
| $NbCr_2$ | 298.8 | 180.5 | 55.5 | 5.493 | 6.193 | 5.648 | 5.629 |

**Table S2**. Theoretical fracture toughness according to Griffith as computed by using the DFT calculations in [36] for three selected crack planes in the $C15$ $NbCr_2$ Laves phase (in units of $MPa \cdot \sqrt{m}$). No information on the termination was provided in [36].

| **Crack plane** | **{111}** | **{110}** | **{100}** |
|---|---|---|---|
| $K_{IC}$ | 1.06 | 1.15 | 1.08 |

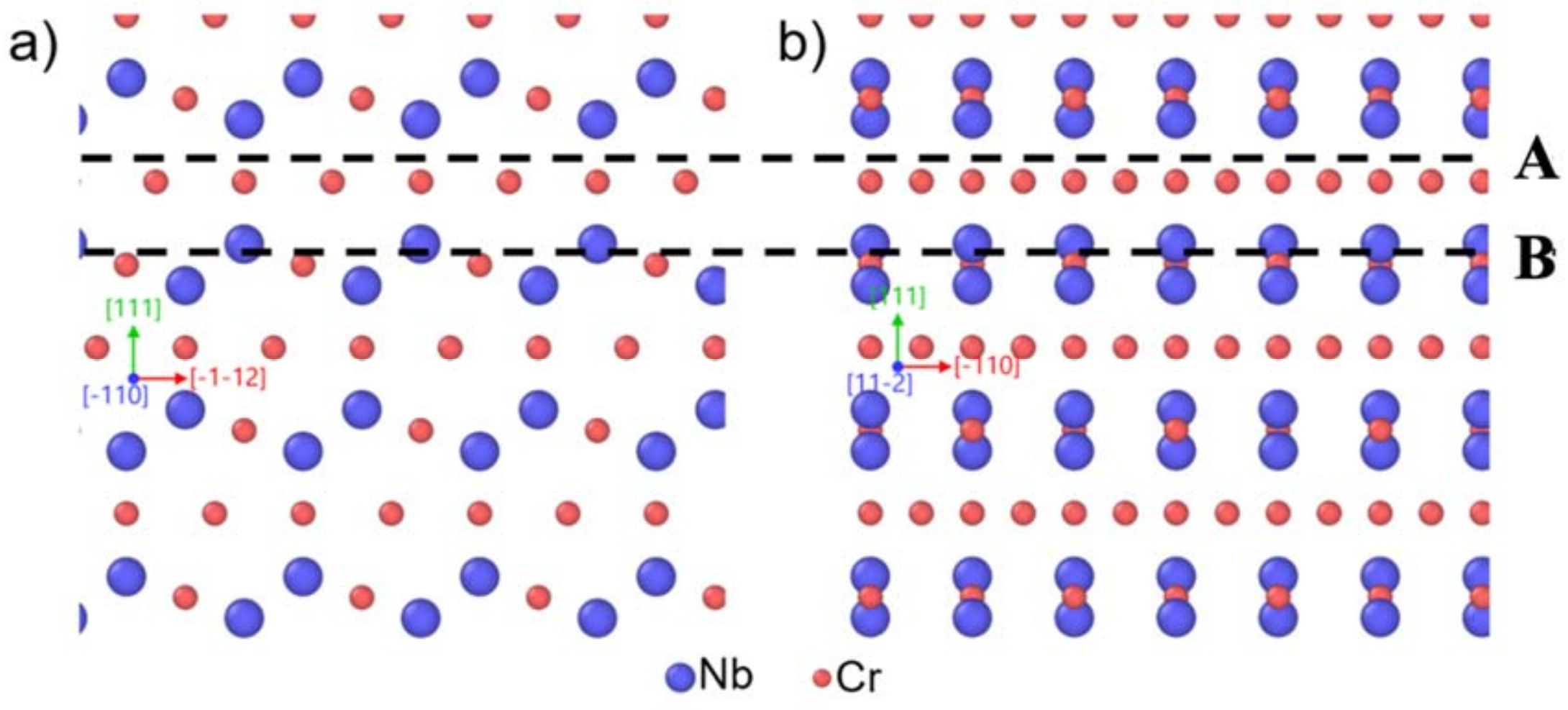

**Figure S7**. Definition of cleavage plane terminations for the (111) plane in the $C15$ Laves phase structure. The crack may run along either of these terminations leading to different crack tip environments for [-110] ($a$) and [11-2] ($b$) out of plane projections.

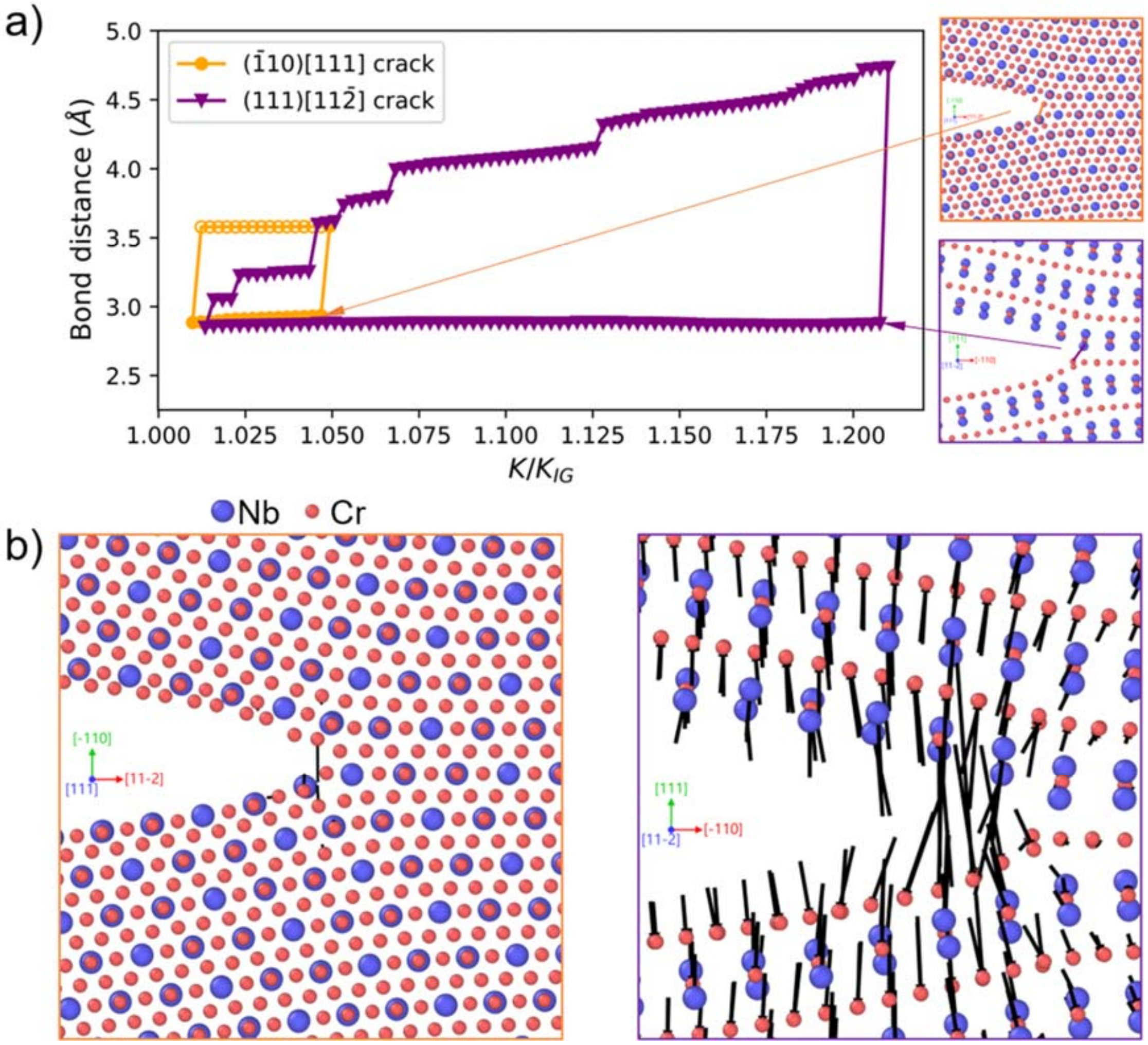

**Figure S8**. a) Variation of the distance between atomic bonds at the crack tip, shown by solid line in the insets, versus $K/K_{IG}$ for (-110)[111] and the B-terminated (111)[11-2] crack systems in our *C*15 $NbCr_2$ model system corresponding to loading and unloading states. b) Atomic positions around the crack tip after the first breaking of atomic bonds. The arrows show the relative displacement of each atom with respect to the previous unbroken state. The displacement magnitudes are magnified 5x.

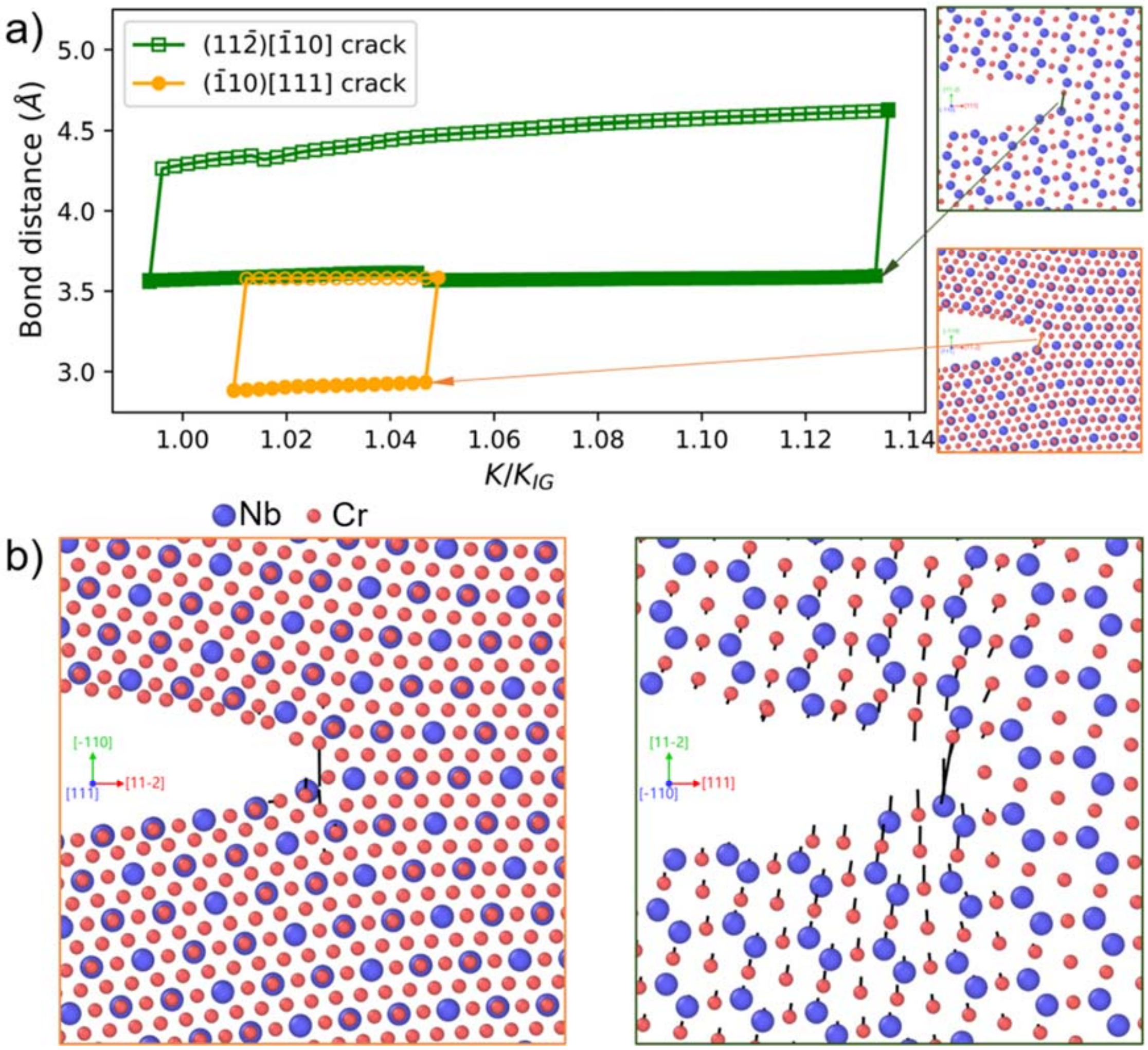

**Figure S9.** a) Variation of the distance between atomic bonds at the crack tip, shown by solid line in the insets, versus $K/K_{IG}$ for (-110)[111] and (11-2)[-110] crack systems in our C15 $NbCr_2$ model system corresponding to loading and unloading states. b) Atomic positions around the crack tip after the first breaking of atomic. The arrows show the relative displacement of each atom with respect to the previous unbroken state. The displacement magnitudes are magnified 5x.